\documentclass[twocolumn,showpacs,preprintnumbers,amsmath,amssymb]{revtex4}
\usepackage{graphicx}
\usepackage{dcolumn}
\usepackage{bm}
\usepackage{here}

\begin{document}


\title{Some Effective Tight-Binding Models for Electrons in DNA Conduction:A Review}

\author{Hiroaki Yamada}
\email{hyamada@uranus.dti.ne.jp}
\homepage{http://www.uranus.dti.ne.jp/~hyamada/}
\affiliation{YPRL, Aoyama 5-7-14-205, Niigata 950-2002, JAPAN}
\author{Kazumoto Iguchi}
\affiliation{KIRL, 70-3 Shinhari, Hari, Anan, Tokushima 774-0003, Japan}


\date{\today}

\begin{abstract}
Quantum transport for DNA conduction has widely studied with 
interest in application as a candidate in making nanowires as well as interest
in the scientific mechanism.
In this paper, we review recent works with concerning the electronic states 
and the conduction/transfer in DNA polymers.  
We have mainly investigated the energy band structure and the correlation effects of
localization property in the two- and three-chain systems (ladder model) 
with long-range correlation as a simple model for electronic property in a double strand 
of DNA by using the tight-binding model.
In addition, we investigated the localization properties of electronic states 
in several actual DNA sequences such as bacteriophages 
of Escherichia coli, human-chromosome 22, compared with those of 
the artificial disordered sequences with correlation. 
 The charge transfer properties for poly(dA)-poly(dT) and poly(dG)-poly(dC) DNA polymers
are also presented in terms of localization lengths within the frameworks of the polaron models
due to the coupling between the charge carriers and the lattice vibrations 
of the double strand of DNA. 
\end{abstract}


\maketitle


\section{Introduction}
Recent interests on semiconducting DNA polymers have been stimulated
by successful demonstrations of the nanoscale fabrication of DNA, where 
current-voltage (I-V) measurements 
for poly(dA)-poly(dT) and poly(dG)-poly(dC) DNA polymers 
have been done \cite{porath00,yoo01,porath04,chakraborty07}.
For such artificial periodic DNA systems, the energy band structure is a useful
starting point in order to interpret the experimental results such as 
semiconductivity and the metal-insulator transition \cite{iguchi04}.
  
On the other hand,   
Tran {\it et al.} measured conductivity along the double helix of
lambda phage DNA ($\lambda-$DNA) at microwave frequencies, using the
lyophilized DNA in and also without a buffer \cite{tran00}. 
The conductivity is strongly temperature
dependent around room temperature with a crossover to a weakly temperature
dependent conductivity at low temperatures \cite{yoo01}. 
Yu and Song showed that the $\lambda-$DNA can be consistently modeled
by considering that electrons may hop through the variable range hopping for conduction 
without invoking additional ionic conduction mechanisms, 
and that electron localization is enhanced by 
strong thermal structural fluctuations in DNA \cite{yu01}.
Indeed, the sequence of base pairs(bp) of the $\lambda-$DNA is inhomogeneous 
as in disordered material systems.

Moreover, charge migration in DNA have been mainly addressed in order to clarify
the mechanism of damage repair which are essential to maintain the integrity of the molecule.
The precise understanding of the DNA-mediated charge migration would
be important on the descriptions of damage recognition
process and protein binding, or on the engineering biological processes \cite{ladik99}. 
The stacked array of DNA bp provides an extended path to a long-range charge transfer 
although the dynamical motions or the energetic sequence-dependent heterogeneities
are expected to reduce the long-range migration. 
Photoexcitation experiments have unveiled that charge excitations can be transfered
between metallointercalators through
 the guanine highest occupied molecular orbitals of the DNA bridge\cite{berlin02}.
The subsequently low-temperature experiments showed that the radiation-induced conductivity 
is related to the mobile charge carriers, migrating within 
frozen water layers surrounding the DNA helix, 
rather than through the base-pair core. 

 Those experiments are summarized as follows:
(i) Band gap reduction of a double strand of DNA,
(ii) Transition from the tunneling hopping to the band hopping,
(iii) Anomalously strong temperature dependence of band gap,
(iv) Highly nonlinear temperature dependence of the DC conductivity,
(v) Low conductivity of DNA with a complicated sequence such as $\lambda-$DNA and
high conductivity of DNA with a simple sequence such as poly(dG)-poly(dC)
and poly(dA)-poly(dT).
These results suggest that the anomalously strong temperature dependence on the
physical quantities is attributed to the "self-organised" 
extrinsic superconductive character of DNA
due to the formation of donors and accepters.
Here we would like to note the following: 
Usually, as in solid state physics, the extrinsic semiconductor is realized 
when external impurities are introduced in pure substrate materials.
Such impurities produce the extrinsic semiconductive nature.
However, in real DNA helices, there are no such impurities from outside;
but there exist already a complicated arrangement of bases of 
adenine (A), guanine(G), cytosine (C) and thymine (T)
inside of the DNA, where among the bases each one of bases may regard 
other bases as impurities.
Hence, a kind of "self-organized" extrinsic semiconductive nature may appear.

However, the electronic transport properties in DNA are still controversial
mainly due to the complexity of the experimental environment and the molecule itself. 
Although theoretical explanations for the phenomena have been tried by
some standard pictures used in solid state physics such as polarons, solitons,
electrons and holes, the situation is still far from unifying the theoretical scheme.

 Each DNA sequence is packed in a chromosome, varying in length from $10^5bp$
 in yeast to $10^9bp$ in human. 
In general, the length of a mutation is relatively short ($10bp$) 
as compared to the length of a gene ($10^3-10^6bp$). 
Because the mutation rate is very law 
the mechanical and thermodynamic characteristics are maintained for the mutation \cite{lin06}.
 In Sect.6, we give a brief discussion about the mutation as the proton transfer
between the normal and tautomer states.

The Watson-Crick(W-C) base-pair sequence is essential for DNA 
to fulfill its function as a
carrier of the genetic code. 
The specific characteristics 
is also used  extensively even in various fields such as anthropology and criminal probe.
  As observed in the power spectrum, the mutual information analysis and the
Zipf analysis of the DNA base sequences such as human chromosome 22(HCh-22), 
the long-range correlation exists in the total sequence, as well as the short-range periodicity.  
In this article, we mainly discuss the relationship between 
the correlation in the DNA base sequences and the electronic transport/transfer.

The electrical transport of DNA is closely related to the density of itinerant $\pi-$electrons 
because of the strong electron-lattice interaction. 
Resistivities of two typical DNA molecules, such as poly(dG)-poly(dC) and $\lambda-$DNA,  
 are calculated. 
 At the half-filling state, the Peierls phase transition takes place 
and the poly(dG)-poly(dC) and poly(dA)-poly(dT) DNA polymers
exhibit a large resistivity. 
When the density of itinerant $\pi-$electrons departs far from the half-filling state, 
the resistivity of poly(dG)-poly(dC) becomes small. 
For the $\lambda-$DNA, there is no Peierls phase transition 
due to the aperiodicity of its base pair arrangement. 
The resistivity of poly(dG)-poly(dC) decreases as the length of the molecular chain is increasing, 
while that of $\lambda-$DNA increases as the length is increasing.

In Sect.3, we introduce the H$\ddot{u}$ckel model of DNA molecules to treat with the 
$\pi-$electrons. 
Moreover, in Sect.4, we give the effective polaron model including the 
electron-lattice coupling dynamics.
In Sect.5,  we present correlation effects of localization in the ladder models and
 the formation of localized polarons (Holstein's polarons or solitons) 
 due to the coupling between the charge carriers and the lattice vibrations of
the double strand of DNA. 
These polarons in DNA act as donors and acceptors and
exhibit an extrinsic semiconductor character of DNA. 
The results are discussed in the context of experimental observations.


The present review article essentially follows the line in Refs.
\cite{iguchi01,iguchi03,iguchi04,yamada04,yamada05,yamada07}.
In particular,  it is written with a view of the localization and/or delocalization problem 
in the quasi-one-dimensional tight-binding models with disorder. 
 In appendices we give some calculations and explanations related to main text.

\section{Correlated DNA Sequences}
As is well known, DNA has the specific binding properties, 
i.e., only A-T and G -C pairs are possible, where the bases of nucleotide are A, T, G, C.
The backbones of the bases, sugar and phosphate groups,
 ensure the mechanical stability of the double helix and protect the base pairs. 
Since the phosphate groups are negatively charged, 
the topology of the duplex is conserved only if it is immersed into an aqueous
solution containing counterions such as $Na^+$ and $Mg^+$ that neutralize the phosphate groups.

The clustering of similar nucleotides can be clarified by studying the properties of the cluster size
distributions on the various real DNA sequences, 
ranging from the viral to higher eukaryotic sequences.
 It is shown that the distribution function $P(S)$ about the number $S$ of the consecutive C-G 
or A-T clusters becomes  $P(S) \propto \exp\{-\alpha S\}$  \cite{holste03,isohata03}. 
The values of the scaling exponent $\alpha$ of CG are much larger than $\alpha$ of AT. 
The maximum value of the A-T cluster size is found to be much larger than that of the C-G cluster size, 
which implies the existence of large A-T clusters.

Moreover, it has been found that the base sequences of various genes exhibit a long-range correlation,
characterized by the power spectrum $S(f) \sim f^{-\beta}$ ($0.1 < \beta < 0.8$) in the low
frequency limit ($f<<1$) \cite{carpena02}. 
As was observed in the power spectrum, 
the mutual information analysis and the Zipf analysis of the DNA base sequences 
such as the HCh-22, 
the long-range structural correlation exists in the total sequence as well as the short-range periodicity. 
The eukaryote's DNA sequence has an apparently periodic repetition in terms of the gene duplication. 
The correlation length in the base sequence of genes 
changes from the early eukaryote to the late eukaryote as a result of evolutionary process. 
It is found that the long-range correlation tends to manifest in the power spectrum of the total sequence
rather than in the power spectra of the exon and intron parts, separately\cite{grosse02}.

 On the other hand, 
in a DNA molecule, the charge carriers move along a double-helix formed 
by two complementary sequences of four basic nucleotides: A, T, G, and C. 
A conduction band would form, if the DNA texts would exhibit some periodicity. 
The electrical resistance of the DNA molecule strongly fluctuates 
even if a single nucleotide in a long sequence is replaced (or removed).
Quantum transport through the DNA molecule is also strongly affected by the correlations.
The localization property of a single-chain  disordered systems 
with long-range correlation has been also extensively studied. 
The correlated disorder can lead to delocalized states 
within some special energy windows in the thermodynamic limit.

Accordingly, it  is very interesting to compare the localization nature 
of the electronic states in the real DNA sequence
with those in the artificial disordered sequence with long-range correlation. 
Recently, Krokhin {\it et al.} have reported that much longer localization
length has been observed in the exon regions than in the intron regions 
for practically all the allowed energies and for all randomly selected DNA sequences\cite{krokhin09}. 
Through the statistical correlations in the nucleotide sequence, 
they suggest that the persistent difference of the localization property 
is related to the qualitatively different informations stored by exons and introns.

In the Sect.5, we numerically give localization nature of the electronic states 
in some real DNA sequences such as bacteriophages of Escherichia coli (E.coli) and HCh-22, 
and so on in ladder models \cite{yamada04,yamada04c}. 
We also investigate the correlation effect on the localization property 
of the one-electronic states in the disordered ladder models with 
a long-range structural correlation that is generated by the modified Bernoulli map. 
 Obviously, the correlation in the DNA sequences affects 
not only the electronic conduction  
but also the twist vibration of the backbones.
However, such effects can be approximately ignored in the  scope of this article.

\section{HOMO-LUMO Gaps}
Each compositional unit of a DNA polymer is complex  
although the DNA polymers can be regarded as quasi-one-dimensional systems.
(See Fig.1.)
In this section, we give a brief review on the H$\ddot{u}$ckel theory in order to analyze
the relationship between electronic structures of the separate nucleotide groups and
of their infinite periodic chains.
\begin{figure}[!h]
\begin{center}
\includegraphics[scale=.4]{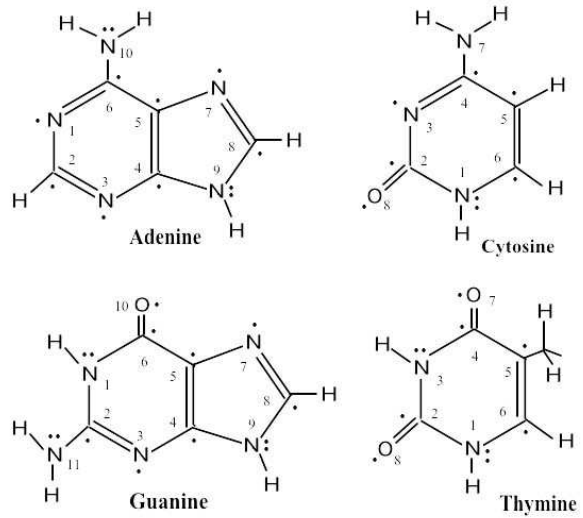}
 \caption{
 The single bases of A, G, C, T. 
Here dots ($\bullet$) mean $\pi$-electrons.
}
\end{center}
\end{figure}
 Ladik tentatively concluded that electrons
hop mainly between the bases along the helical axis of DNA, so that it is enough
to take into account the overlap integrals between the $\pi-$orbitals of the adjacent
base pairs in the DNA duplexes\cite{ladik99}. 
These studies have revealed that {\it DNA polymers
are insulators\/} with an extremely large band gap $E_g$ (about 10-16 eV) 
and narrow widths of the valence and conduction bands (about 0.3-0.8 eV). 
This is a consequence of the orbital mixing between 
the highest occupied molecular orbital(HOMO) 
and the lowest unoccupied molecular orbital (LUMO) 
in the base groups of nucleotide in the DNA. 
 We can control the charge injection into DNA and the conduction in DNA,  
 by adjusting the HOMO-LUMO of the polymer and the Fermi energy of electric leads,
 respectively. 

Brumel {\it et al.} found that semiconduction in DNA is very important for such systems 
since the activation energies of  nucleotides are lower than those of nucleosides \cite{burmel69}. 
This was the first suggestion that propagation along the sugar and 
phosphate groups play a significant role
in the transport properties of DNA as well as propagation along the base-stacking of the nucleotide
groups in the center of the DNA molecule.

We revisit the problem of the electronic properties of individual molecules of DNA, 
in order to know electron transport in the double strand of DNA 
as a mother material for the single and double strand of DNA, 
by taking into account only $\pi-$electrons in the system\cite{iguchi04}. 
To do so, we review the theory of $\pi-$electrons in DNA, 
using the H$\ddot{u}$ckel approximation for $\pi-$electrons 
in both the sugar-phosphate backbone chain 
and the $\pi-$stacking of the nitrogenous bases of nucleotide.

\subsection{H$\ddot{u}$ckel Model}
Applying the basic knowledge of quantum chemistry to biomolecules of nitrogenous bases, 
we can find the total number of  $\pi-$orbitals and $\pi-$electrons in the bases.  
 It is summarized in Table 1.

\begin{table}
\begin{center}
\begin{tabular}{ccccccc}  
\hline \hline
{\em Base} & A & G & C & T & S & Ph \\ \hline
$|N|$ & 5 & 5 & 3 & 5 & -& - \\  
$|C|$ & 5 & 5 & 4 & 2 & 5& 0 \\ 
$|O|$ & 0 & 1 & 1 & 2 & 4& 4 \\ 
$|P|$ & - & - & - & - & 0 & 1 \\ \hline
{\em Total} & 10 & 11 & 8 & 9 & 9& 5 \\ \hline
{\em $\pi$-orbitals} & 10 &11 & 8 & 8 & 4& 5 \\
{\em $\pi$-electrons} & 12 & 14 & 10 &10 & 8& 8 \\ 
\hline \hline
\end{tabular}
\end{center}
\caption{ 
The total numbers $|N|$, $|C|$, $|O|$ and $|P|$
of N, C, O and P atoms and
of $\pi$-orbitals and $\pi$-electrons in the bases of A, G, C and T, 
and the sugar(S) and phosphate(Ph) groups, respectively.
Here there are 12, 14, 10 and 10 $\pi$-electrons for the 10, 11, 8, 8 $\pi$-orbitals
in the A, G, C and T base molecules, respectively, 
while there are 8 and 8 $\pi$-electrons for the 4 and 5 $\pi$-orbitals
in the sugar and phosphate groups, respectively.
Here we note that the carbon atom of $CH_3$ in the T molecule does not have 
any $\pi$-electron since it forms the $sp^3-$hybrid orbitals. 
This provides 8 $\pi$-electrons for the T molecule.
}
\label{table:1}
\end{table}

Let us consider the famous H$\ddot{u}$ckel model in quantum chemistry
\cite{burmel69,fukui76,huckel32,hoffman52,nagata65,burdett84}. 
This model concerns only the $\pi-$orbitals in the system.
In this context, this theory is closely related to the so-called tight-binding model in
solid state physics, which concerns very localized orbitals at atomic sites such as the
Wannier's wavefunction. 
Therefore, this approach has been extensively applied
to many polymer systems such as polyacetylene with a great success.
In the so-called H$\ddot{u}$ckel approximation we adopt the orthogonality
condition for the overlap integrals
\begin{eqnarray}
S_{rr} = 1,\ S_{rs} = 0\ (r \ne s),  
\end{eqnarray}                 
where $S_{rs}$ denotes overlap integrals between atomic orbitals at $r$th and $s$th sites. 
  We assume the special form of
the resonance integrals
\begin{eqnarray}
h_{rs} = \frac{1}{2}KS_{rs}(h_{rr}+h_{ss}),  
\end{eqnarray}
with $K = 1.75$ and the overlap integrals $S_{rs}$ ($r=s$) are not necessary to be diagonal,
and otherwise.
  Usually the on-site ($r=s$) resonance integrals are called the Coulomb integrals
denoted by $\alpha_{r}$, while the off-site ($r \neq s$) integrals are called the resonance
integrals denoted by $\beta_{rs}$ such that
$\alpha_{r} = h_{rr},\ \beta_{rs} = h_{rs}\ (r \ne s)$.
Here we would like to emphasize the following. In the sense of the H$\ddot{u}$ckel theory,
the parameters are taken empirically. This means that the parameters are adjustable
and feasible to give consistent results with the experimental results or the ab initio
calculation results. Therefore, the exact values of the parameters are neither so
important nor should be taken so seriously in this framework. This is because once
one can obtain much more precise values for the H$\ddot{u}$ckel parameters, one can provide
the more plausible results from the H$\ddot{u}$ckel theory.
Although many efforts of the {\it ab initio} calculations have been done for DNA
systems, unfortunately at this moment there seem to be very few
first principle calculations for such parameters in the DNA systems to fill out this
gap. Nevertheless, we must assign some values for the H$\ddot{u}$ckel parameters in
order to calculate the electronic properties of DNA in the framework of the H$\ddot{u}$ckel
theory. So, we look back to the original method about time
when the H$\ddot{u}$ckel theory was invented.

For this purpose to use the standard H$\ddot{u}$ckel theory, let us adopt some simple
formulae for the H$\ddot{u}$ckel parameters, which are defined as follows: Let $X$ and $Y$ be
two different atoms. Denote by $\alpha_X$ the Coulomb integral at the $X$ atom and by
$\beta_{XY}$ the resonance integral between the $X$ and $Y$ atoms:
\begin{eqnarray}
\alpha_{X} = \alpha + a_{X} \beta,                  
\beta_{XY} = l_{XY}\beta. 
\end{eqnarray}            
            
Here $a_X$ and $\ell_{XY}$ are the empirical parameters that are supposed to be adopted
from experimental data. And the parameters $\alpha$ and $\beta$ are important. These can be
thought of the fundamental parameters in our problem of biopolymers. Conventionally
we take $\alpha$ as the Coulomb integral for the $2p_x-$orbital of carbon and $\beta$ as
the resonance integral between the $2p_x-$orbitals of carbon, such that
$\alpha = \alpha_{C} \equiv 0,\ \beta = \beta_{CC} \equiv 1. $
This means that the energy level of a carbon atom is taken as the zero level, and
the energy is measured in units of the resonance integral between carbon atoms.
We note that the empirical values obtained from experiments are usually given
by
\begin{eqnarray}
\alpha \approx - 6.30 \sim -6.61 eV,\
\beta \approx - 2.93 \sim -2.95 eV.                     
\end{eqnarray}
Since the biomolecules consist of the atoms C, N, O, P,
let us find the plausible values of the H\"{u}ckel parameters for them
to apply the H\"{u}ckel model 
to biomolecules of DNA.
The Pauling's electronegativity for 
carbon (C), nitrogen (N), oxygen (O) and phosphorus (P)
are the following:
\begin{eqnarray}
\chi_{C} \approx 2.55,\ 
\chi_{N} \approx 3.0,\ 
\chi_{O} \approx 3.5,\ 
\chi_{P} \approx 2.1.                          
\end{eqnarray}
Then we can defined the H\"{u}ckel parameters
for carbon, using the Sandorfy's formula\cite{sandorfy49}, 
Mulliken's formula\cite{mulliken49} and
Streitwieser's formula\cite{streitwieser61}, 
we find 
$\alpha_{\dot{C}} = \alpha,$ 
$\alpha_{\dot{N}} =  \alpha + 0.7\beta,$
$\alpha_{\dot{O}} =  \alpha + 1.53\beta$, 
$\alpha_{\dot{P}} =  \alpha - 0.724\beta$.    
and  we obtain
$\alpha_{\ddot{O}} =  \alpha + 2.53\beta,$
$\alpha_{\ddot{N}} =  \alpha + 0.276\beta$. 
and 
$\beta_{\dot{C}-\dot{C}} = \beta$,
$\beta_{\dot{C}=\dot{C}} = 1.1\beta$, 
$\beta_{\dot{C}-\dot{N}} = \beta_{\dot{C}-\ddot{N}} = 0.8\beta$,
$\beta_{\dot{C}=\dot{N}} = 1.1\beta$,
$\beta_{\dot{C}-\dot{O}} = 0.9\beta$, 
$\beta_{\dot{C}=\dot{O}} = 1.7\beta$. 
Here $\alpha_{\dot{X}}$($\alpha_{\ddot{X}}$) denotes the Coulomb integral
of atomic state $X$ with one(two) electron(s) occupied.
See appendix A for the formula.
And if one can get more accurate values from the ab initio calculations, then 
we can always replace the Coulomb integrals by the new set of values.

\subsection{HOMO-LUMO of Biomolecules}
For example, considering the topology of the hopping of electrons on the $\pi$-orbitals,
we now find the following H\"{u}ckel matrices 
${\bf H}_{A}$ for A:
\begin{widetext}
\begin{eqnarray}
{\bf H}_{A} = \nonumber 
\left(
\begin{array}{cccccccccc}
\alpha_{\dot{N}} & \beta_{\dot{C}-\dot{N}} & 0 & 0 & 0 & \beta_{\dot{C}=\dot{N}} & 0 & 0 & 0 & 0\\
\beta_{\dot{C}-\dot{N}} & \alpha_{\dot{C}} & \beta_{\dot{C}=\dot{N}} & 0 & 0 & 0 & 0 & 0 & 0 & 0\\
0 & \beta_{\dot{C}=\dot{N}} & \alpha_{\dot{N}} & \beta_{\dot{C}-\dot{N}} & 0 & 0 & 0 & 0 & 0 & 0\\
0 & 0 & \beta_{\dot{C}-\dot{N}} & \alpha_{\dot{C}} & \beta_{\dot{C}=\dot{C}} & 0 & 0 & 0 & \beta_{\dot{C}-\ddot{N}} & 0\\ 
0 & 0 & 0 & \beta_{\dot{C}=\dot{C}} & \alpha_{\dot{C}} & \beta_{\dot{C}-\dot{C}} & \beta_{\dot{C}-\dot{N}} & 0 & 0 & 0\\
\beta_{\dot{C}=\dot{N}} & 0 & 0 & 0 & \beta_{\dot{C}-\dot{C}} & \alpha_{\dot{C}} & 0 & 0 & 0 & \beta_{\dot{C}-\ddot{N}}\\
0 & 0 & 0 & 0 & \beta_{\dot{C}-\dot{N}} & 0 & \alpha_{\dot{N}} & \beta_{\dot{C}=\dot{N}} & 0 & 0\\
0 & 0 & 0 & 0 & 0 & 0 & \beta_{\dot{C}=\dot{N}} & \alpha_{\dot{C}} & \beta_{\dot{C}-\ddot{N}} & 0\\
0 & 0 & 0 & \beta_{\dot{C}-\ddot{N}} & 0 & 0 & 0 & \beta_{\dot{C}-\ddot{N}} & \alpha_{\ddot{N}} & 0\\
0 & 0 & 0 & 0 & 0 & \beta_{\dot{C}-\ddot{N}} & 0 & 0 & 0 & \alpha_{\ddot{N}}\\
\end{array}
\right),      \nonumber                                 
\end{eqnarray}
\end{widetext}
Pauli's exclusion principle tells us that each state with an energy level is occupied by
a pair of electrons with spin up and down. So, $\pi-$electrons occupy the energy levels
in the spectrum from the bottom at low temperature. Since the lower energy levels
with one half of the total number of $\pi-$electrons can be occupied by the $\pi-$electrons,
there appears an energy separation between the occupied and the unoccupied states,
which is called the energy gap. 

The energy levels of the HOMOs and LUMOs are given by $\varepsilon_{H} = 0.888$, 
$\varepsilon_{L} = -0.789$ for A, where the energy is measured in units of $\beta$. 
 We can do the same calculation for G,C,T, respectively.
Defining the energy gap between the LUMO and HOMO, 
$\Delta \varepsilon = \varepsilon_{L} - \varepsilon_{H}$.
Then we obtain the result as, 
$\Delta \varepsilon_{A} =1.677$, $\Delta \varepsilon_{G}=1.555$,
$\Delta \varepsilon_{C}=1.535$, $\Delta \varepsilon_{T}=1.713$.
 (See Fig.2(a).) 
The  total energy $E_{tot} = 2 \sum_{j=occ. states} \varepsilon_{j}$ 
of the $\pi-$electrons of A, G, C, T are 
$E_{tot}(A) = 20.74$, $E_{tot}(G) = 26.47$, $E_{tot}(C) = 19.05$, $E_{tot}(T) = 21.14$, respectively.
Therefore, 
$$E_{tot}(C) > E_{tot}(A) > E_{tot}(T) > E_{tot}(G). $$
This shows that since the lower
the ground state energy the more stable the system, the most stable molecule is G
while the most unstable molecule is C.

\begin{figure}[!h]
\begin{center}
\includegraphics[scale=.3]{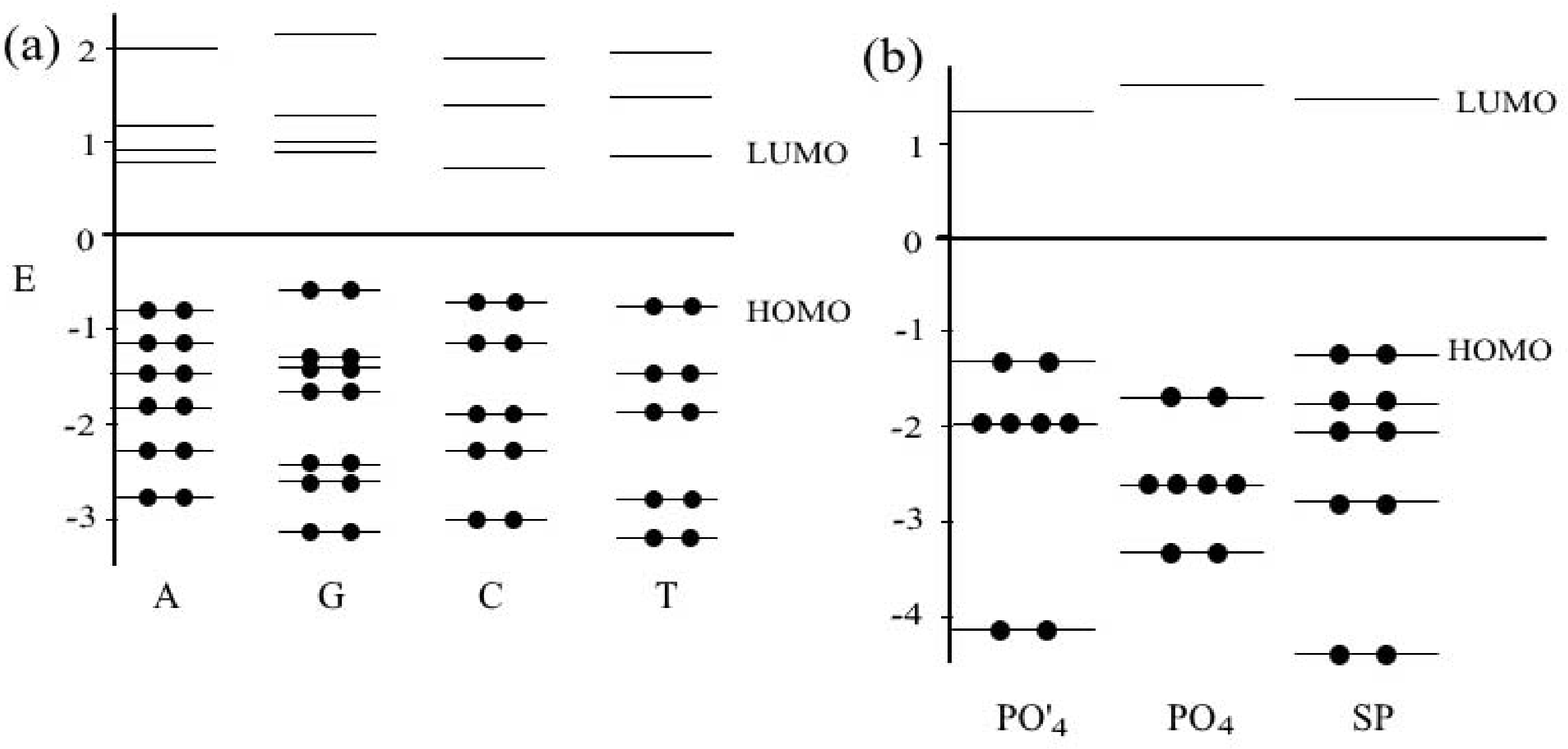}
 \caption{
(a) The spectrum of the $\pi$-orbitals of A, G, C, T. 
(b) The spectrum of the $\pi$-orbitals of a sugar-phosphate group. 
PO$_{4}$ stands for the phosphate group where the electron hopping between
the oxygen sites is taken into account, while
PO'$_{4}$ stands for the the phosphate group where the electron
hopping between the oxygen sites is forbidden.
SP stands for the sugar-phosphate group where
the electron hopping between
the oxygen sites is taken into account.
The energies are measured in units of $|\beta|$.
Here dots ($\bullet$) mean $\pi$-electrons
and the level with four dots means the double degeneracy
of the level.
}
\end{center}
\end{figure}

HOMOs and LUMOs of sugar and phosphate may be required 
when we consider charge conduction of the DNA polymer 
 because the sugar and phosphate group constituting the backbones of DNA polymer 
 also have $\pi-$electrons. 
 We give the results for the single sugar-phosphate in Fig.2(b).
Moreover, the results based on the H$\ddot{u}$ckel approximation have been extended to 
the single nucleotide systems such as A, G, C, T with the single sugar-phosphate
group, and
the system of a single strand of DNA with an infinite repetition of a nucleotide
group such as A, G, C and T, respectively.
See Ref.(\cite{iguchi04}) for the more details.
This reorganization, which is difficult to
calculate due to the complexity of the combined system, may lead to 
a smaller HOMO-LUMO gap and wider band widths than for the bare molecule.

When the system of the nucleotide bases such as A, G, C, T exists as an individual
molecule, there is always an energy gap between the LUMO and HOMO
states, where the order of the gap is about several eV.
This means that the nucleotides of A, G, C, T have the semiconducting character
in its nature.
If the $\pi-$stacking of the base is perfect, then
there are two channels for $\pi-$electron hopping: One is the channel through
the base stacking and the other is the one through the backbone chain of the
sugar-phosphate. In this limit, the localized states of the original nucleotide
bases become extended such that the levels form the energy bands. 
 
Thus, we believe that our H$\ddot{u}$ckel approach in this paper fills out the gap between
the simple approach of mathematical models for tight-binding calculations and the
approach of the quantum chemical models for ab initio calculations from the first
principle, where in the former we assume one orbital with one electron at one
nucleotide while in the latter we include all electrons and atoms in the system.

\section{Effective Polaron Models}
First principle methods are powerful enough to understand the basic electronic states
of the molecules.
On the other hand, the complementary model-based Hamiltonian approach is effective 
for understanding those of polymers as well.
In this section, we give some effective 1D models and the basic properties
of polymers such as trans-polyacetylene and DNA polymers.
 In the last section, we treated HOMOs and LUMOs and the occupation of orbitals 
by the $\pi-$electrons with spin.  
  Hereafter, we omit spin of electrons 
when we consider the one-electron problems, for the sake of simplicity in notation.

\subsection{Tight-binding Models for Polymers}
It is known that the electronic properties of 
 planar conjugated systems are dominated by $\pi-$systems with one orbital per site. 
The electronic Hamiltonian 
for $2p_z$ orbitals through a tight-binding model with the nearest neighbors interactions only:
 \begin{eqnarray}
H_{el} & = & \sum_{n} E_{n}C_{n}^{\dagger}C_{n} 
- \sum_{n}V_{nn+1}(C_{n}^{\dagger}C_{n+1}+C_{n+1}C_{n}^{\dagger}),
\end{eqnarray}
 where $C_{n}$ and $C_{n}^{\dagger}$ are creation and annihilation
operators of an electron at the site $n$. 
The matrix elements are obtained from the extended H$\ddot{u}$ckel theory
as given in the previous section:
\begin{eqnarray}
 E_n &= & -\alpha_{n} \\
 V_{nm} &=& (K/2)(\alpha_{n} +\alpha_{m}) S_{nm}, 
\end{eqnarray}
where $\alpha_{n}$ is ionization energy (Coulomb integral) of the $n$th $2p_z$ orbital,
and $S_{nm}$ is overlap integral (resonance integral) between the $n$th 
and $m$th orbitals centered at 
neighbour sites given in Sect.3.
 We usually treat the copolymer as a $\pi-$system with one 
orbital per site and represent the electronic Hamiltonian for the $2p_z$ 
carbon and nitrogen orbitals through a tight-binding model with 
nearest neighbor interactions.
  DNA polymers are also believed to form an effectively one-dimensional
molecular wire, which is highly promising for diverse applications. 
Basically, the carriers mainly propagate along the aromatic $\pi-\pi$ stacking 
of the strands (the interstrand
coupling being much smaller), so that the one-dimensional tight-binding chain model can be
a good starting point to minimally describe a DNA wire\cite{iguchi94}.

 Bruinsma {\it et al.} also introduced the effective tight-binding model that 
describes the site energies of 
 a carrier located at the $n$th molecule as, 
\begin{widetext}
\begin{eqnarray}
H =\sum_n E_n C_n^\dagger C_n 
-\sum_n t_{DNA} \cos(\theta_{n,n+1}(t)) (C_n^\dagger C_{n+1} +C_{n+1}^\dagger C_{n}), 
\end{eqnarray}
\end{widetext}
where $C_n^\dagger $($C_n$)  creation(annihilation) operator of the carrier at the site $n$
\cite{bruinsma00,yu01,roche03}.
The carrier site energies $E_n$ are chosen according
to the ionization potentials of respective DNA bases as,
$\epsilon_A=8.24eV$, $\epsilon_T=9.14eV$, $\epsilon_C=8.87eV$, and 
$\epsilon_G=7.75eV$, while the hopping integral, simulating the
$\pi-\pi$ stacking between adjacent nucleotides, is taken as
$t_{DNA}=0.1-0.4eV$. 
$\theta_{n,n+1}$ denotes the relative twist angle deviated from equilibrium
position between the $n$th and $(n+1)$th molecules due to temperature $T$.
Then we can estimate the order of the hopping integral.
Let $\theta_{n,n+1}$ is an independent random variable that follows 
Gaussian distribution with $\langle \theta_{n,n+1}\rangle=0$ and  
$\langle \theta_{n,n+1}^2\rangle=k_BT/I\omega_I^2$, where $I$ is the reduced moment
of inertia for the relative rotation of the two adjacent bases and $\omega_I$
is the oscillator frequency of the mode ($I\omega_I^2=250K$).
Therefore the fluctuation of the Hopping term is about 
$t_{DNA}\langle \cos(\theta_{n,n+1})\rangle \sim t_{DNA}(1-\langle \theta_{n,n+1}^2\rangle/2) \sim 0.16eV$
 at room temperature, 
 which is much smaller than that ($\sim 1.4 eV$) in the diagonal part.
Accordingly, as a simple approximation, we can deal with the diagonal fluctuation 
 with keeping the hopping integral constant. 

 Roche investigated such a model for poly(GC) DNA polymer  and $\lambda-$DNA, with the on-site disorder
arising from the differences in ionization potentials of the base pairs, and with the bond
disorder $t_{DNA} \cos(\theta_{n,n+1}(t)) $ related to the random 
twisting fluctuations of the nearest-neighbor bases
along the strand\cite{roche03}. 
While for poly(GC) the effect of disorder does not appear to be very dramatic,
the situation changes when considering $\lambda-$DNA.

Actually, modern development of physico-chemical 
experimental techniques enables us to measure
directly the DNA electrical transport phenomena even in single molecules. 
Moreover, several groups have recently
performed numerical investigations of localization properties 
of the DNA electronic states based on the realistic DNA sequences.

\subsection{Twist Polaron Models for DNA Polymers}
In this subsection we give an effective 1D model with realistic parameters for the
electronic conduction of the poly(dA)-poly(dT) and poly(dG)-poly(dC)
DNA polymers.

Specially, a distinctive feature of biological polymers is 
a complicated composition of their elementary subunits, 
and an apparent ability of their structures to support long-living nonlinear excitations. 
In their polaron-like model, Hennig and coworkers studied the electron breather propagation along DNA
homopolynucleotide duplexes, i.e. in both poly(dG)-poly(dC) and poly(dA)-poly(dT) DNA polymers, 
and,
for this purpose, they estimated the electron-vibration coupling strength in DNA, 
using semiempirical quantum chemistry\cite{starikov02,hennig02,palmero04}.
 Chang {\it et al.} have also considered a possible mechanism to explain the phenomena 
of DNA charge transfer.
The charge coupling with DNA structural deformations can create a polaron 
and thus promote a localized state. 
As a result, the moving electron breather may contribute 
to the highly efficient long-range conductivity\cite{chang04}. 
Recent experiments seem to support the polaron mechanism
for the electronic transport in DNA polymers.

The Hamiltonian for the electronic part in the DNA model
is given by
 \begin{eqnarray}
H_{el}(t) & = & \sum_{n}E_{n}(t)C_{n}^{\dagger}C_{n} 
- \sum_{n}V_{nn+1}(t)(C_{n}^{\dagger}C_{n+1}+C_{n+1}C_{n}^{\dagger}),
\end{eqnarray}
 where $C_{n}$ and $C_{n}^{\dagger}$ are creation and annihilation
operators of an electron at the site $n$. 
The on-site energies $E_{n}(t)$
are represented by
\begin{eqnarray}
E_{n}(t) & = & E_{0}+kr_{n}(t),
\end{eqnarray}
 where $E_{0}$ is a constant and $r_{n}$ denotes the structural fluctuation
caused by the coupling with the transversal Watson-Crick H-bonding stretching
vibrations. 
The schematic illustration is given in Fig.3.
  The transfer integral $V_{nn+1}$ depends on the three-dimensional
 distance $d_{nn+1}$ between adjacent stacked base pairs, labeled by $n$ and
$n+1$, along each strand.  And it is expressed as 
\begin{eqnarray}
V_{nn+1}(t) & = & V_{0}(1-\alpha d_{nn+1}(t))\,.
\end{eqnarray}
Parameters $k$ and $\alpha$ describe the strengths of
interaction between the electronic and vibrational variables, respectively.
The 3D displacements $d_{nn+1}$  also give rise to variation of the distances
between the neighboring bases along each strand.  
The  first order Taylor expansion around the equilibrium positions is given by
\begin{eqnarray}
d_{nn+1}(t) & = & \frac{R_{0}}{\ell_{0}}(1-\cos\theta_{0})(r_{n}(t)+r_{n+1}(t)).
\end{eqnarray}
 $R_{0}$ represents the equilibrium radius of the helix, $\theta_{0}$
is the equilibrium double-helical twist angle between base pairs, and $\ell_{0}$
the equilibrium distance between bases along one strand given by
\begin{eqnarray}
\ell_{0} & = & (a^{2}+4R_{0}^{2}\sin^{2}(\theta_{0}/2))^{1/2},
\end{eqnarray}
 with $a$ being the distance between the neighboring base pairs in the
direction of the helix axis. 
We adopt realistic values of the parameters obtained from
the semi-empirical quantum-chemical calculations. (See table 2.)
  Further, we consider $\{ r_{n}\}$ as
independent random variables generated by a uniform distribution
with the width ($r_{n}\in[-W,W]$). 
Accordingly, fluctuations in both the
 on-site energies and the off-diagonal parts in the Hamiltonian (1)
are mutually correlated because they are generated by the same
random sequence $r_{n}$. (See Fig.2(c).) 
The typical value of $W$ is $W=0.1 [\AA]$, which approximately corresponds
 to the variance in the hydrogen bond lengths in the
Watson-Crick base pairs, as seen in the X-ray diffraction experiments
\cite{xray84}.

A quasi-continuum spectrum is possessed of a wealth of dynamical modes. 
In principle, each of these can influence the DNA charge transfer/transport.
But, since there are more or less active modes,
it is possible to take the whole manifold of the DNA molecular vibrations into two parts: 
vibrations which are most active and  a  stochastic bath  consisting of all the other ones. 
Computer simulations have pinpointed that  the dynamical disorder
is crucially significant for the DNA transfer/transport.
Several attempts to formulate stochastical models for the interplay of the former and
the latter have already appeared in the literature(see, for examples).
In this paper, we will deal with the polaronlike model of Hennig and coworkers 
as described in the works\cite{hennig02,palmero04}, 
where charge+breather propagation along DNA homopolynucleotide duplexes
[i.e. in the both poly(dG)-poly(dC) and poly(dA)-poly(dT) DNA
polymers] has been studied.

\begin{table}
\begin{center}
\begin{tabular}{cc}  
\hline \hline
 {\em parameter} & {\em value} \\ \hline
$k_{AT}$ & 0.778917 $eV \AA^{-1}$ \\
$\alpha_{AT}$  & 0.053835 $\AA^{-1}$ \\ 
$k_{GC}$ & -0.090325 $eV \AA^{-1}$ \\ 
$\alpha_{GC}$ & 0.383333  $\AA^{-1}$   \\
\hline \hline
\end{tabular}
\end{center}
\caption{ 
Basic parameters for DNA molecules. 
The subscripts, $AT$ and $GC$, for $k$ and $\alpha$ denote 
for ones of the poly(dA)-poly(dT) and poly(dG)-poly(dC) DNA polymers, respectively. 
The other parameters are $E_0 = 0.1eV$,  $V_0=0.1eV$,  $a=3.4 \AA$,  $R_0=10 \AA$ and 
$\theta_0=36^{\circ}$.
}
\end{table}

\begin{figure}[!h]
\begin{center}
\includegraphics[scale=.6]{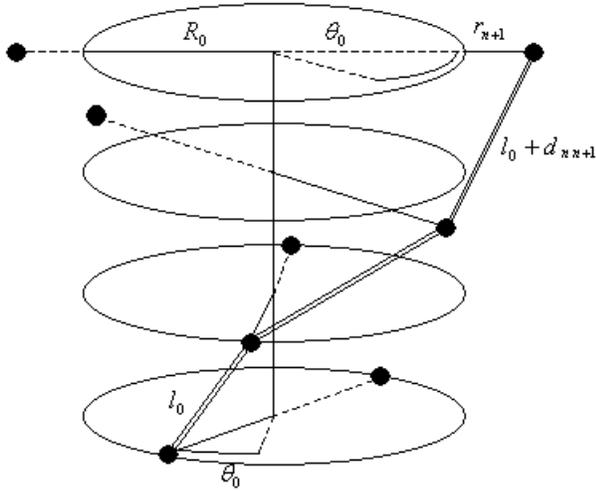}
 \caption{
Sketch of the structure of the DNA model. The bases are represented
by bullets and the geometrical parameters $R_0, \ell_0, \theta_0, r_{n+1}$
 and $d_{nn+1}$
are indicated.
}
\end{center}
\end{figure}

\begin{figure}[!h]
\begin{center}
\includegraphics[scale=.6]{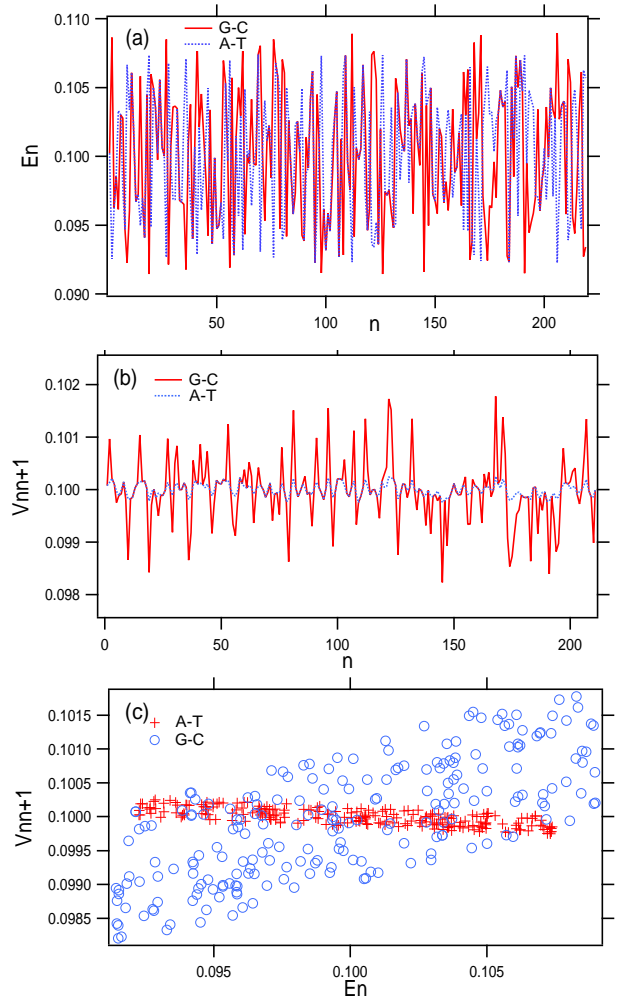}
 \caption{(a)The on-site energy $E_neV$ and (b)transfer integral
 $V_{nn+1}eV$ as a function of the base pair
site $n$. The parametric plot $E_n$ versus $V_{nn+1}$ is shown in (c).
$W=0.1$ and the other parameters are given in text. 
The unit of the energy and the spatial length are $eV$ and 
the number of nucleotide base pair($bp$), respectively, throughout the present paper. 
}
\end{center}
\end{figure}

\subsection{Localization due to Static Disorder}
In this subsection, 
we assume that the fluctuations of $\{ r_n \}$ are frozen (quenched disorder) and 
independent for different links;
that is, we investigate localization properties of 
the poly(dA)-poly(dT) and poly(dG)-poly(dC) DNA polymers\cite{yamada01}.

The Schr$\ddot{o}$dinger equation $H_{el}|\Phi \rangle=E|\Phi \rangle $ 
is written in the transfer matrix form,
\begin{eqnarray}  
 \left(
\begin{array}{c}
\phi_{n+1} \\
\phi_{n} \\ 
\end{array}
\right)
=
\left(
\begin{array}{cc}
\frac{E-E_{n} }{V_{nn+1}} & -\frac{V_{nn-1}}{V_{nn+1}}  \\
 1 & 0 
\end{array}
\right)
\left(
\begin{array}{c}
\phi_{n} \\
\phi_{n-1} \\ 
\end{array}
\right), 
\label{transfer}
\end{eqnarray}  
\noindent
 where $\phi_{n}$ is the
 amplitude of the electronic wavefunction
 $|\Phi\rangle =\sum_n \phi_n  |n\rangle $, 
 where $|n \rangle= C_n^\dagger|0 \rangle$  at  the base pair site $n$.
$|0\rangle$ is the Fermi vacuum.
 We use the localization length $\xi$ and/or 
Lyapunov exponent $\gamma $ calculated by the mapping (\ref{transfer}) 
in order to characterize the exponential 
localization of the wave function. 
   Originally the Lyapunov exponent (the inverse localization length) is
defined in the thermodynamic limit ($N\rightarrow\infty$).
However, here we use the following definition of  the Lyapunov exponent  for the electronic wave function
with a large system size $N$ \cite{crisanti93,yamada01}.
\begin{eqnarray}
\gamma(E,N)=\xi^{-1}(E,N)=\frac{\ln(|\phi_{N}|^{2}+|\phi_{N-1}|^{2})}{2N}.
\end{eqnarray}
 We use the appropriate initial conditions $\phi_{0}=\phi_{1}=1$, 
and  for large $N(>>\xi)$  the localization length and
the Lyapunov exponent are independent of the boundary condition. 
The energy-dependent transmission coefficient $T(E,N)$ of the system between 
metallic electrodes is given as $T(E,N)=\exp (-2\gamma N)$ and is related to 
Landauer resistance via $\varrho =(1-T)/T$ 
in units of the quantum resistance $h/2e^2( \sim 13[k\Omega m])$ \cite{lifshits88,imry02}.


\begin{figure}[!h]
\begin{center}
\includegraphics[scale=.5]{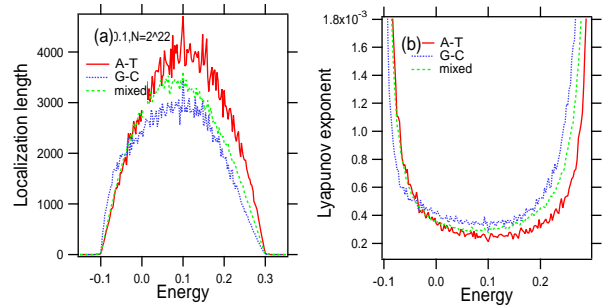}
 \caption{
Comparison (a)localization length and (b)Lyapunov exponent in
poly(dG)-poly(dC), poly(dA)-poly(dT) and the mixed DNA polymers.
$W=0.1$, $\theta_0=36^\circ$ and $N=2^{22}$.
}
\end{center}
\end{figure}

\begin{figure}[!h]
\begin{center}
\includegraphics[scale=.45]{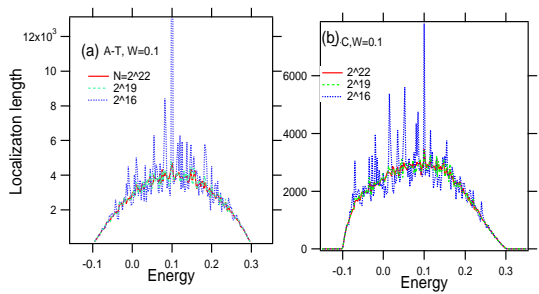}
 \caption{
The localization length as a function of
the energy for several system sizes
 $N(=2^{16}, 2^{19},2^{22})$ in the (a)poly(dA)-poly(dT) DNA
polymers, (b)poly(dG)-poly(dC)  DNA polymers.
$W=0.1$ and $\theta_0=36^\circ$. }
\end{center}
\end{figure}

We have numerically investigated the localization properties of
electronic states in the adiabatic polaron model of poly(dG)-poly(dC) and poly(dA)-poly(dT)
DNA polymers with the realistic parameters obtained using the
semi-empirical  quantum-chemical calculations.

We compare the localization properties
of the poly(dG)-poly(dC), poly(dA)-poly(dT) DNA polymers and the mixed cases. 
Figure 5(a) and (b) show the localization length and
the Lyapunov exponent in the three types of polymers with $W=0.1$. 
In low energy regions, the localization length in the poly(dG)-poly(dC) DNA
polymer is larger than that in the poly(dA)-poly(dT) DNA polymer.
The system size dependence of the localization
length is given in Fig.6 in relation to the resonance energy.
Indeed, the localization length of the DNA polymers is larger than $\xi>2000[bp]$ 
in almost all the energy bands  for all models; 
it is much larger than the system size of the oligomer used in the experiments. 
As is seen in Fig.6,  the smaller the size of the system
the more complex the resonance peaks become.

\subsection{Quantum Diffusion in the Fluctuating Environment}
In this subsection we numerically investigate
quantum diffusion of electrons in the Hennig
model of poly(dG)-poly(dC) and poly(dA)-poly(dT) with a
dynamical disorder\cite{yamada07}. 
We assume that the diffusion is caused by a colored
noise associated with the stochastic dynamics of
distances r(t) between two Watson-Crick base pair partners:
 $ \langle r(t)r(t^{'}) \rangle = r_{0} \exp(-|t-t^{'}|/\tau)$.
 These fluctuations can be regarded as a stochastic process at high temperature,
with phonon modes being randomly excited\cite{kats02,benjamin09}. 
In the model, the characteristic decay time $\tau$ of correlation
can control the spread of the electronic wavepacket. 
Interestingly, the white-noise limit  $\tau \to 0$ can in effect correspond
to a sort of motional narrowing regime(see, examples),
because we find that such a regime causes
ballistic propagation of the wavepacket through homogeneous DNA duplexes. 
Still, in the adiabatic limit  $\tau \to \infty$,
DNA electronic states should be strongly affected by localization.
The amplitude  $r_0$ of random fluctuations within the base pairs 
(the fluctuation in the distance between two bases in a base pair) 
and the correlation time  $\tau$,
are very critical parameters for the diffusive properties of the wavepackets. 
It is interesting to find the ballistic behavior in the white 
noise limit vs the localization in the adiabatic limit,
since there is a number of experimental works
observing ballistic conductance of DNA in water solutions,
which is also temperature-independent. 
Zalinge et al. have tried to explain the latter effect, using a kind
of acoustic phonon motions in DNA duplexes, 
which seems to be plausible, but not the only possible physical
reason\cite{zalinge06,mandal06}. 
 We will propose an alternative explanation
for the observed temperature-independent conductance,
based upon our numerical results.

Generally, in the case of quantum diffusion
the temporal evolution of the electronic state vector $|\Phi  \rangle$ is described by the 
time-dependent Schr$\ddot{o}$dinger equation 
$i\hbar \frac{ \partial |\Phi  \rangle }{\partial t} = H_{el}(t) |\Phi  \rangle$,
which then becomes 
\begin{eqnarray}
i \hbar_{eff} \frac{ \partial \phi_n }{\partial t} & = & 
E_n(t) \phi_n -V_{nn+1}(t) \phi_{n+1} -V_{n-1n}(t) \phi_{n-1} ,  
\end{eqnarray}
where $\phi_n = \langle n|\Phi  \rangle$ and the effective Planck constant $\hbar_{eff}=0.53$.
We redefined the scaled dimensionless variables
 $E_n(t)$ and $V_{n n+1}(t)$ in Eq.(1) such that
 $\frac{E_n}{V_0} \to E_n$,  $\frac{V_{nn+1}}{V_0} \to V_{nn+1}$.  
We used mainly the 4th order Runge-Kutta-Gill method 
in the numerical simulation for the time evolution with
 time step $\delta t=0.01$.  
In some cases, we confirmed ourselves that the accuracy in the obtained results  
is in accord with the one in the results that are gained by the help of 
the 6th order symplectic integrator that is the higher order unitary integration.

\begin{figure}[h]
\begin{center}
\includegraphics[scale=0.5,clip]{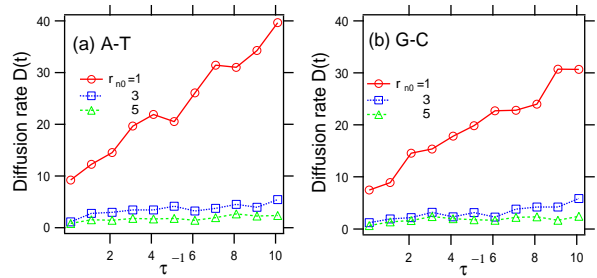}
 \caption{Diffusion rate $D(t)$ as a function of $\tau^{-1}$ 
for several fluctuation strengths $r_{n0}=1, 3, 5$
at (a)A-T model and (b)G-C model, respectively
}
\end{center}
\end{figure}

\begin{figure}[h]
\begin{center}
\includegraphics[scale=0.5,clip]{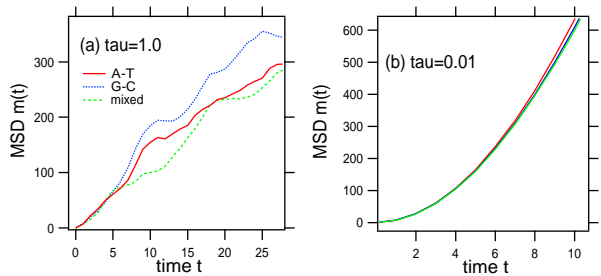}
 \caption{Short-time behavior of $m(t)$ of A-T, G-C, and mixed models 
with $r_{n0}=1.0$ at (a)$\tau=1$ and 
 (b)$\tau=0.01$.
}
\end{center}
\end{figure}
It has been demonstrated that the motional narrowing affects the
localization in the poly(dG)-poly(dC) and poly(dA)-poly(dT) DNA polymers.
In either case of the model DNA polymers, 
the temperature-dependence becomes virtually suppressed when the motion of the 
wave packet is characterized by the ballistic propagation. 

We have also investigated the temporal diffusion rate in almost all the diffusive ranges.  (See Fig.7.)
We found that the diffusion rate of the  A-T model is larger than that of 
the G-C model at comparatively low temperatures. 
Interestingly enough, for relatively high temperatures in the diffusive range of the wavepacket motion 
the difference between the two DNA systems gets smaller.

Figure 8 shows the short-time behavior in the cases 
of $\tau=1$ and $\tau=0.01$ depicted on a larger time scale. 
It follows that in the short-time behavior 
the  $(dG)_{15}-(dC)_{15}$ case is more diffusive 
than its $(dA)_{15}-(dT)_{15}$ counterpart within the range 
from which the spread of the wavepacket is 
$\surd m  \sim 15$.

We used periodic sequences, which means that $E_0$ and $V_0$ are constant values
as the static parts of the on-site and hopping terms 
for poly(dG)-poly(dC) and poly(dA)-poly(dT) DNA polymers, respectively. 
This includes the mixed model as well.
Then the motional narrowing for dynamical disorder 
makes time-evolution of the wavepacket ballistic.  
However, we remark that the motional narrowing strongly localizes the 
wavepacket if we use a disordered sequence for the static parts of 
$E_n$ and/or $V_{nn+1}$.

\subsection{Quantum Diffusion Coupled with Vibrational Modes}
Let us denote $r_{n}(t)$ by the stretching vibration of  W-C bonds at the $n$ site as before.
The external harmonic perturbation is equivalent to the coupling with quantum linear oscillators or 
with phonon modes in solid state physics \cite{yamada99,yamada04a}.
We replace the disordered fluctuation $r_n(t)$ with the harmonic time-dependent one such as
\begin{eqnarray}
\label{PAM}
   r_n(t)=r_{n0} \sum_{i=1}^M \epsilon_i \cos (\omega_i t +\theta_n^{(i)} ),
\end{eqnarray}
where $M$ and $\epsilon_i$ are the number of the frequency component and 
the strength of the perturbation, respectively.
The initial phases  $\{ \theta_n^{(i)} \} \in [0,2\pi] $ at each site $n$ are randomly chosen.
In the numerical calculation of this section, we take 
 $\epsilon_i=\frac{\epsilon}{\surd M}$, for simplicity, 
and take incommensurate numbers of $\omega_i \sim O(1)$ as the frequency set.
In the limit of $M \to \infty$, the time-dependent perturbation approaches
the stochastic fluctuation as discussed in the last subsection.
 In particular, we can regard  the approximation as an electronic system
coupled with highly excited quantum harmonic oscillators. 
One of the advantage of this model is that although the number of autonomous modes 
is limited due to the computer power, 
we can freely control the number $M$ of frequency components in the 
harmonic perturbation.
This is also a simple model to investigate electronic diffusion coupled with lattice vibrations.

In Fig.9(a), the MSD is shown for the poly(dG)-poly(dC), the poly(dA)-poly(dT) 
and the  mixed DNA polymers. 
We find that all the cases exhibit the normal diffusion of electron without any 
stochastic perturbation. 
As shown in Fig.9(b), the diffusion rate decreases as the 
perturbation strength $\epsilon$ increases. 
As a result the coupled motion of charges and the lattice breathers connected with the
localized structural vibrations may contribute to the highly
efficient long-range conductivity. (See appendix D.)

\begin{figure} [h]
\begin{center}
\includegraphics[scale=0.5,clip]{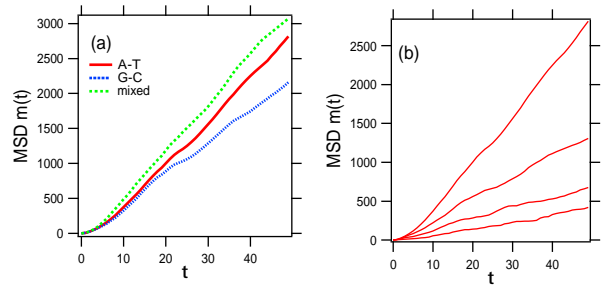}
 \caption{ (a) MSD $m(t)$ in A-T, G-C, and mixed models perturbed by 
$M=1,\epsilon=0.5$.
(b) MSD $m(t)$ of A-T model perturbed by $M=1$ for 
$\epsilon=0.5,0.7,1.0,2.0$ from above. We set $r_{n0}=1$.
}
\end{center}
\end{figure}


\section{Tight-binding Models for the Ladder Systems}
In this section, to investigate the energy band structure for periodic sequences and 
the localization properties for the correlated disordered sequences,
we introduce the ladder models of DNA polymers.
Although in Refs.\cite{iguchi01,yamada04,yamada04c}, 
we assumed the ladder structure consisting of sugar and phosphate groups,
 the model can be applicable to the DNA-like substances such as 
the ladder structure of bases pairs without sugar-phosphate backbones.
When we apply the model for the sugar and phosphate chains, 
the chain $A$ and $B$ are constructed by the repetition 
of the sugar-phosphate sites, and the inter-chain hopping $V_n$
at the sugar sites come from the nucleotide base-pairs, i.e., $A-T$ or $G-C$ pairs.   
(See Fig.10(a).)

\begin{figure}[!h]
\begin{center}
\includegraphics[scale=0.35]{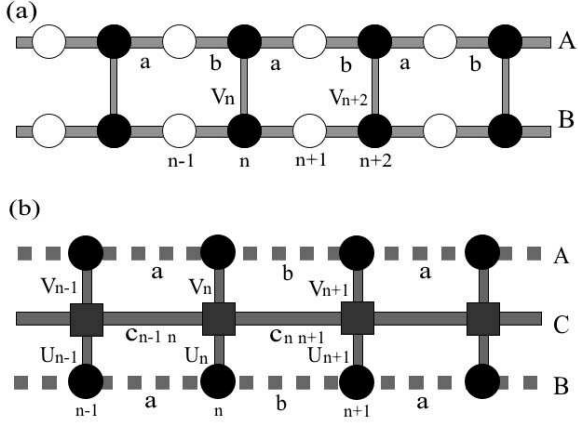}
  \caption{
Models of the double strand of DNA.
(a) the two-chain model, (b) the three-chain model we adapted in the main text.
}
\end{center}
\end{figure}

\subsection{The Ladder Models and the Dispersions}
Since there is only one $\pi-$orbital per site, 
there are totally $2N$ $\pi-$orbitals in the ladder model of the DNA double chain. 
Let us denote by $\phi_n^A$($\phi_n^B$) the $\pi-$orbital at site n in the chain $A$($B$).
By superposition of the $\pi-$orbitals, 
   the Schr$\ddot{o}$dinger equation
$\hat{H}|\Psi\rangle = E|\Psi\rangle$ becomes,
\begin{eqnarray*}  
A_{n+1,n} \phi^{A}_{n+1}+ A_{n,n-1} \phi^{A}_{n-1} + A_{n,n} \phi^{A}_{n} 
+ V_{n} \phi^{B}_{n}  = E\phi^{A}_{n}, \\ 
B_{n+1,n} \phi^{B}_{n+1}+ B_{n,n-1} \phi^{B}_{n-1} + B_{n,n} \phi^{B}_{n} 
+ V_{n} \phi^{A}_{n}  = E\phi^{B}_{n},  
\end{eqnarray*}  
\noindent
where $A_{n+1,n}$ ($B_{n+1,n}$) means the hopping integral 
between the $n$th and $(n+1)$th sites and $A_{n,n}$ ($B_{n,n}$) 
the on-site energy at site $n$ in chain $A$
($B$), and $V_{n}$ is the hopping integral 
from chain $A (B)$ to chain $B (A)$ at site $n$,
respectively.  
 Furthermore it can be rewritten in the matrix form, 
$  \Phi_{n+1} = M_n  \Phi_{n} $, 
where $\Phi_{n}=(\phi^{A}_{n},\phi^{A}_{n-1},\phi^{B}_{n},\phi^{B}_{n-1})^\dagger $.
  The transfer explicit matrix is given in appendix B.
  Generally speaking, we would like to investigate the asymptotic 
behavior ($N \to \infty$) of the products of the matrices 
$M_{d=2}(N)=\Pi_{k=1}^N M_k = M_N M_{N-1} \cdots M_1$. 

We consider the band structure for the periodic case by setting
 $A_{n+1,n} = B_{n+1,n}=a(b)$ at odd (even) site $n$
 and $V_n=v(0)$ at odd(even) sites, 
 $A_{nn}=B_{nn}=\alpha$($A_{nn}=B_{nn}=\beta$) for odd(even) site $n$, 
 for simplicity. 
 
   A simple way to solve the above equations is to use the Bloch theorem:
\begin{eqnarray}    
\phi^{A}_{n+2} = e^{i2ks}\phi^{A}_{n},  \phi^{B}_{n+2} = e^{2iks}\phi^{B}_{n},
\end{eqnarray} 
where $s$ is the length between the adjacent base groups 
and is assumed to be equivalent to the length between the adjacent sugar-phosphate groups
and the wavenumber $k$ is defined as $-\frac{\pi}{2s} \le k \le \frac{\pi}{2s}$.
 (We take $s=1$ in the following calculations.)
 The Schr$\ddot{o}$dinger equation becomes 
 \begin{widetext}
\begin{eqnarray} 
  \Phi_{n+1} &=& M  \Phi_{n}, \\
M &= &
\left( 
\begin{array}{cccc}
E - \beta & -(a+be^{ik})  & 0 & 0\\
-(a+be^{-ik})  & E - \alpha & 0 & -v \\
0 & 0 & E-\beta & -(a+be^{ik})  \\
0 & -v & -(a+be^{-ik}) & E-\alpha 
\end{array}
\right),
\end{eqnarray} 
\end{widetext}
where $\Phi_{n} =(\phi_{2n+1}^A, \phi_{2n}^A, \phi_{2n+1}^B,\phi_{2n}^B)^t$.
Denote the determinant of $M$ by $D(M)$.
Solving the equation $D(M)=0$ for $E$, 
 we obtain the energy dispersion of the system. 
 Here we show the simple case of $\alpha=\beta=0$, 
 \begin{eqnarray}
 E_{+}^{(\pm)}(k) = \frac{1}{2} \left(  v \pm \sqrt{ v^2+4(a^2+b^2) + 8ab\cos 2k } \right),  \\
 E_{-}^{(\pm)}(k) = \frac{1}{2} \left( -v \pm \sqrt{ v^2+4(a^2+b^2) + 8ab\cos 2k } \right). 
\end{eqnarray}
$E_{\pm}^{(+)}(k)$ ( $E_{\pm}^{(-)}(k)$ ) stand for the upper(lower) bands in 
channel $\pm$, respectively\cite{iguchi01}.
The lowest and upper middle (the lower middle and highest) bands correspond to bonding (antibonding)
 states between the adjoint orbitals in the interchains.
  The more detail of the energy band for general case is given in appendix C.
In Fig.11(a) the energy band structure is given with varying the interchain hopping $v$.
Figure 11(b) shows the cross-section view at $v=1$.
 There is a band gap $E_g(v)$ at the center in between the lower and upper middle bands in the
 spectrum for the whole range of $v$ when $\alpha=\beta=0$.  
\begin{eqnarray}
E_g(v) &=&  E_{-}^{(+)}(\pi/2)-E_{+}^{(-)}(\pi/2)= \sqrt{4(a-b)^2 + v^2 }-v. 
\end{eqnarray} 
The other band gap $\Delta_g(v)$,  
\begin{eqnarray}
\Delta_g(v) &=& v +\frac{1}{2} \left( \sqrt{4(a-b)^2 + v^2} - \sqrt{4(a+b)^2 + v^2} \right), 
\end{eqnarray} 
appears in between the lowest and the lower middle bands (the upper middle and highest bands) 
when $v > v_c \equiv 2ab/\surd a^2+b^2$ (otherwise, it is negative and therefore semimetallic).
There is a transition from semiconductor to
semimetal as the $\pi-$electron hopping between the nitrogenous bases of nucleotide is
increased.

The two-chain model can be easily extended to the three-chain case.
 (See Fig.10(b).)
$C_{n+1,n}$  means the hopping integral 
between the $n$th and $(n+1)$th sites and $C_{n,n}$  
the on-site energy at site $n$ in chain $C$, and $V_{n}$ 
and $U_n$ are the hopping integral between the chains.
  The Schr$\ddot{o}$dinger equation
$\hat{H}|\Psi\rangle = E|\Psi\rangle$ becomes,
\begin{widetext}
\begin{eqnarray}  
A_{n+1,n} \phi^{A}_{n+1}+ A_{n,n-1} \phi^{A}_{n-1} + A_{n,n} \phi^{A}_{n} 
+ V_{n} \phi^{C}_{n}  = E\phi^{A}_{n}, \nonumber \\ 
C_{n+1,n} \phi^{C}_{n+1}+ C_{n,n-1} \phi^{C}_{n-1} + C_{n,n} \phi^{C}_{n} 
+ V_{n} \phi^{A}_{n} + U_{n} \phi^{B}_{n}   = E\phi^{C}_{n}, \\ 
B_{n+1,n} \phi^{B}_{n+1}+ B_{n,n-1} \phi^{B}_{n-1} + B_{n,n} \phi^{B}_{n} 
+ U_{n} \phi^{C}_{n}  = E\phi^{B}_{n},  \nonumber
\end{eqnarray} 
\end{widetext} 
\noindent
It can be also rewritten in the matrix form,
$\Phi_{n+1}= M_{d=3}(n) \Phi_n$
where $\Phi_{n}=(\phi^{A}_{n+1},\phi^{A}_{n},\phi^{C}_{n+1},\phi^{C}_{n},\phi^{B}_{n+1},\phi^{B}_{n})^t$ and 
the explicit transfer matrix is given in the appendix C.

For simplicity, we set $A_{n+1,n} = B_{n+1,n}=a$ and $C_{n+1,n}=c$ at site $n$,  and $V_n=U_n$. 
We can analytically obtain the energy band structure for 
the three chain model by using the Bloch theorem.
\begin{eqnarray}  
\phi_{n+1}^A = e^{iks}\phi_{n}^A,\ \ 
\phi_{n+1}^C = e^{iks}\phi_{n}^C, 
\phi_{n+1}^B = e^{iks}\phi_{n}^B.
\end{eqnarray} 


\begin{figure}[!h]
\begin{center}
\includegraphics[scale=.7]{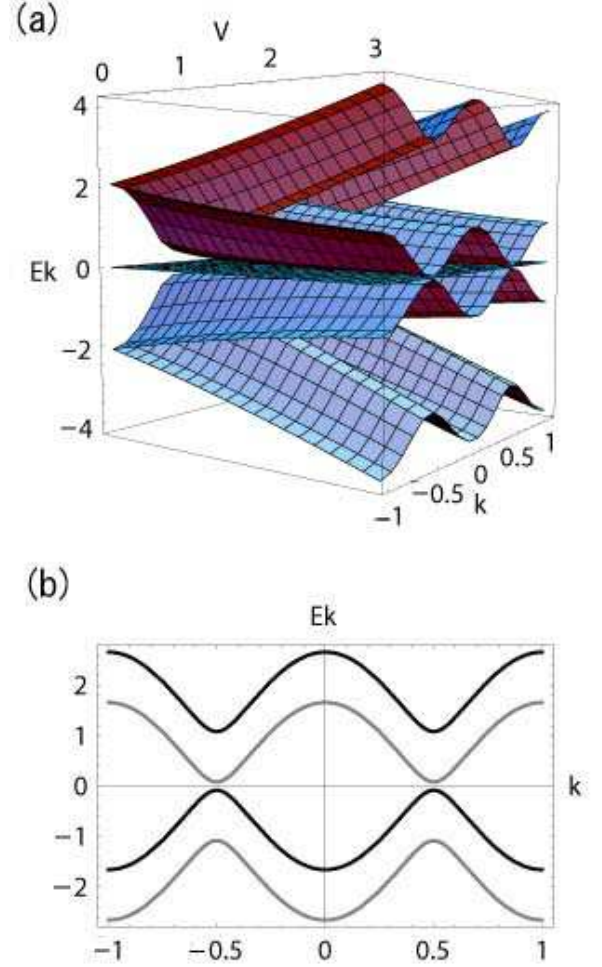}
\caption{
Energy bands of the decorated ladder model.
(a) The energy bands as a function of $V$.
(b) The snap shot of the energy bands when $V = 1$.
$k$ means the wave vector in units of $\frac{\pi}{s}$ such that $-1.0 \le k \le 1.0$,
$V$ means the $\pi$-electron hopping integral between
the inter chain sites, and
$E$ means the energy in units of $V = 1$.
Here we have taken the values 
$\alpha= 0$, $\beta= 0$, $a=0.9,b=1.2$.
}
\end{center}
\end{figure}

\subsection{Localization in the Ladder Models}
As was discussed in Sect. 2 and Sect. 3, 
the sequences of the realistic DNA polymers are not periodic and accompany a variety of disorder.
 Generally, the randomness makes the electronic states localized 
 because of the quasi-one-dimensional system 
 and affects electronic conduction and optical properties, and so on. 
In the present section, we investigate the correlation effect on the localization
properties of the one electronic states in the disordered, two-chain (ladder) and
three-chain models with a long-range structural correlation 
as a simple model for the electronic property in the DNA\cite{yamada05}. 
The relationship between the correlation length in the DNA sequences and 
the evolutionary process is suggested. 
Moreover, it is interesting if the localization property is related to the evolutionary process. 

Figure 12 shows the DOS as a function of energy for the binary disordered systems.
The sequence of the interchain hopping  $V_n$ takes an alternative value $W_{GC}$ or $W_{AT}$. 
We find that some gaps observed in the periodic cases 
close due to the disorder corresponding to the base-pair sequences.

Figure 13(a) shows the energy dependence of the Lyapunov exponents 
($\gamma_1$ and $\gamma_2$) for some cases in the asymmetric modified Bernoulli system 
characterized by the bifurcation parameters $B_0$ and $B_1$ controlling the correlation. 
(See appendix G for modified Bernoulli system.)
They are named as follows: 
Case(i):$B_0 = 1.0, B_1 = 1.0$, Case(ii):$B_0 = 1.0, B_1 = 1.9$ and Case(iii):$B_0 = 1.7, B_1 = 1.9$. 
Apparently, the case(i) is more localized than the cases(ii) and (iii) in the vicinity of 
the band center $|E| < 1$. 
The comparison between the case(ii) and the case(iii) shows the effect of asymmetry 
of the map on the localization. 
The ratio of the G-C pairs $R_{GC} \sim 0.2$ for the case(ii), 
$R_{GC} \sim 0.47$ for the case(iii). 
In the energy regime $|E| > 1$, the Lyapunov exponent $\gamma_2$ in the case(ii) is smaller than 
the one in the case(iii) in spite of the same correlation strength $B_1$. 
As a result, we find that in the DNA ladder model, 
correlation and asymmetry enhance the localization length $\xi (\equiv 1/\gamma_2$ ) 
of the electronic states around $|E| < 1$, 
although the largest Lyapunov exponents $\gamma_1$ do not change effectively at all.
Figs. 13(b), (c) and (d) show the nonnegative Lyapunov exponents in the real DNA sequences 
of (b) HCh-22, (c) bacteriophages of E. Coli and (d) histon proteins. 
In the case of HCh-22, we used two sequences with $N = 105$, extracted 
from the original large DNA sequences. 
The result shows that  the Lyapunov exponents do not depend the detail 
of the difference in the sequences of HCh-22. 
Although the weak long-range correlation has been observed in HCh-22 
as mentioned in the introduction, 
it does not affect on the property of  localization. 
The sequences we used are almost symmetric ($R_{GC} \sim 0.5$).

\begin{figure}[!h]
\begin{center}
\includegraphics[scale=.6]{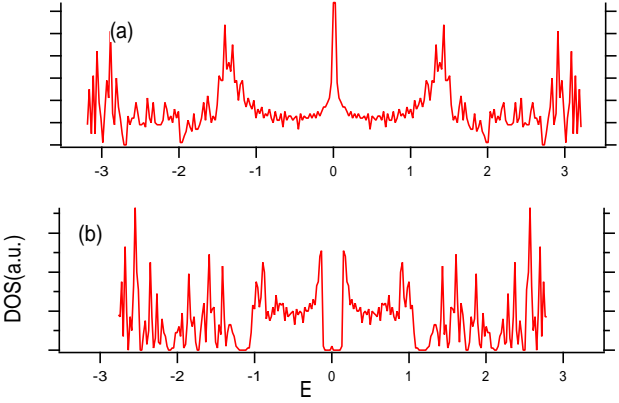}
\caption{DOS as a function of energy for the binary disordered cases 
in the two-chain model.
 (a)$W_{AT}=2.0, W_{GC}=1.0, a=1.0, b=1.0$.
(b)$W_{AT}=2.0, W_{GC}=1.0, a=1.0, b=0.5$. 
The on-site energy is set st $A_{nn}=B_{nn}=0$.
}
\end{center}
\end{figure}

\begin{figure}[!h]
\includegraphics[scale=.5]{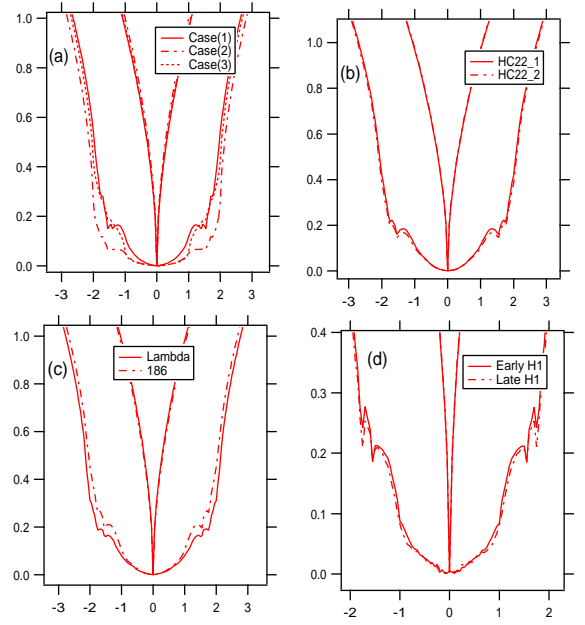}
  \caption{
Lyapunov exponents ($\gamma_1, \gamma_2$) as a function of energy in the ladder model. 
(a) Modified Bernoulli model, where 
cases(1):$B_0 = 1.0, B_1 = 1.0$, Case(2):$B_0 = 1.0, B_1 = 1.9$ and 
Case(3):$B_0 = 1.7, B_1 = 1.9$ are shown. 
(b)HC-22. 
(c) bacteriophages of E. coli (phage-$\lambda$, phage-186). 
(d) early histone H1 and late histone H1. 
The on-site energy is set at $A_{nn} = B_{nn} = 0$, $a =-1.0$,
$b =-0.5$. The size of the sequence is $N = 10^5$ for (a), $N = 10^5$ for (b), 
$N = 48510$ for the phage-$\lambda$ in (c), $N = 30624$ for the phage-186 in
(c), $N = 787$ for the early histone H1 in (d), and $N = 1182$ for 
the late histone H1 in (d).
}
\end{figure}

\begin{figure}[!h]
\includegraphics[scale=.5]{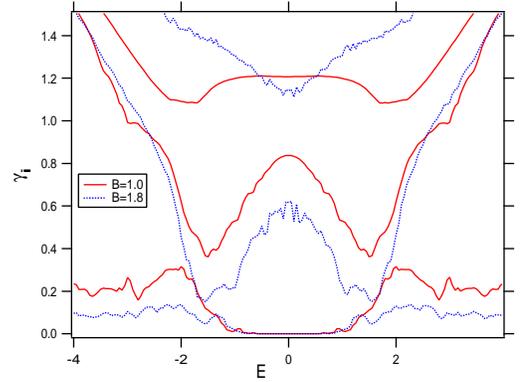}
  \caption{Lyapunov exponents $\gamma_i (i=1,2,3)$
  as a function of energy in the correlated three-chain model.
The parameters $W_{AT}=2.0, W_{AT-AT}=1.0, a=1.0, b=0.5$.
The on-site energy is set at $A_{nn}=B_{nn}=C_{nn}=0$.
}
\end{figure}

In addition, in the numerical calculation 
we set the on-site energy as $A_{nn}=B_{nn}=C_{nn}=0$ for simplicity.
 The sequence $\{ C_{nn+1} \}$ can be also generated by corresponding to the 
base-pairs sequence.
 The localization properties in the simple few-chain models with the on-site
disorder have been extensively investigated \cite{heinrichs02}. 
Figure 14 shows the energy dependence of the Lyapunov exponents 
in the three chain cases. 
 We can observe that all the Lyapunov exponents $\gamma_i(i\leq d) $ are changed 
by the correlation. The least nonnegative Lyapunov exponent $\gamma_3$ is diminished by
the correlation. 
 In particular, it is found
that the correlation of the sequence enhances the localization length defined by
using the least non-negative Lyapunov exponent. 
 We can see that the localization length diverges at the band center $E=0$.

\begin{figure}[!h]
\begin{center}
\includegraphics[scale=.5]{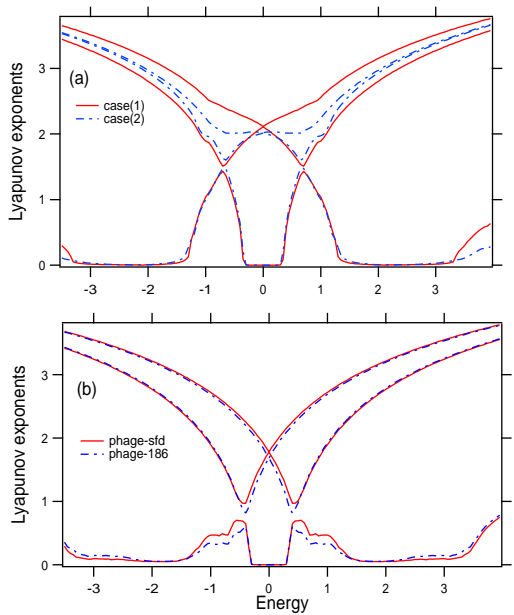}
  \caption{
Lyapunov exponents ($\gamma_1,\gamma_2,\gamma_3$) as a function of energy 
in the three-chain model. 
(a) Modified Bernoulli model, (b) bacteriophages of
E. coli (phage-$fd$, phage-186). The parameters are same as ones in Fig.14 except for on-site
 energies of C-chain (Cn,n = 0).
 }
\end{center}
\end{figure}

In Fig.15(a) the result for asymmetric modified Bernoulli system is shown. 
Apparently, the correlation and/or asymmetry of the sequence 
effect a change in the second and third Lyapunov exponent. 
In contrast, although the global feature of $\gamma_1$ is almost unchanged, 
the local structure of the energy dependence is changed by the change
of $B_0$. 
Fig.15(b) shows the results in phage-fd and phage-186 in the three-chain model. 
It is found that the structure of the energy dependence around $|E| < 2$ 
is different from that in the artificial sequence by the modified Bernoulli map.

The charge transfer efficiency based on mismatch and correlation effect for 
some genomic and synthetic sequences are also investigated \cite{roche04}

\subsection{Polaron Models for a Double Strand of DNA}
The overlap of the electronic orbitals along the stacked
base pairs provides a pathway for charge propagation over 50$\AA$  and 
the reaction rate between electron donors and acceptors does not decay exponentially with distance.
A multiple-step hopping mechanism was recently proposed
to explain the long-range charge transfer behavior in DNA.
In this theory, the single G-C base pair is considered as a hole donor 
due to its low ionization potential if compared to the one on the A-T base pairs.
Besides the multiple-step hopping mechanism, 
polaron motion has also been considered as a possible mechanism
to explain the phenomena of DNA charge transfer.
The charge coupling with the DNA structural deformations may create a polaron 
and cause a localized state.
The polaron behaves as a Brownian particle such that 
it collides with low energy excitations of its environment that acts as a heat bath.
For the most physical systems, the acoustical and optical phonons 
are the main lattice excitations as in the single-chain case given in Appendix D. 

In this section, we present the formation of localized polarons 
due to the coupling between charge carriers and lattice vibrations of a double strand of DNA\cite{iguchi03}. 
These polarons in DNA act as donors and acceptors, which exhibit an extrinsic semiconductor character of DNA. 
The results are discussed in the context of experimental observations.

Let $x_n$ and $y_n$ be configuration angles for rotation of the $n$th base group 
along the course of the backbone chains A and B of DNA ladder, respectively. 
For the sake of simplicity, we assume that all masses $M_n$ of the nucleotide groups and
the bond lengths $d_n$ between $S$ and $B_n$ groups are identical so that $M_n = M$ and
$d_n = d$, giving the moment of inertia $I_n = I_0 = Md^2$. 
In this case, the lattice Hamiltonian $H_{ph}$ can be given by
\begin{widetext}
\begin{eqnarray}
H_{ph}(x_n,y_n) = \sum^N_{n=1} \{ \frac{I_0}{2}( \dot{x}_n^2 + \dot{y}_n^2 ) 
+\frac{A}{2}( x_n^2 + y_n^2 )  
+ \frac{K}{2}( x_{n+1}-x_{n})^2  \nonumber \\
+ \frac{K}{2}( y_{n+1}-y_{n})^2  
+\frac{S}{2}(x_n - y_n)^2  \} 
\end{eqnarray}
\end{widetext}
where A, K and S are the parameters for the rotational energy, the stacking energy
and the bonding energy of the bases, respectively. 
We note here that various 
generalizations and modifications of the model are straightforwardly possible by
adapting a different combination of the interactions and the hopping integrals.
 In the following, we give result only for HOMO. 
We can obtain the result for LUMO in simple replacements.

On the other hand, generalizing the idea of Holstein for the one-dimensional
molecular crystal to our case of the double strand of DNA, 
 the tight bingding Hamiltonian for electrons is given by
 \begin{widetext}
\begin{eqnarray*}
H_{el}^H = \sum_{n=1}^N
\{ -ta_{(n+1)}^{A \dagger} a_{n}^A -t a_{(n-1)}^{A \dagger} a_{n}^A 
+ \epsilon_{H}^A(x_n) a_{n}^{A \dagger} a_{n}^A + v_n(|x_n-y_n|) a_{n}^{A \dagger} a_{n}^B \} \\
+\sum_{n=1}^N
\{ -ta_{(n+1)}^{B \dagger} a_{n}^B -t a_{(n-1)}^{B \dagger} a_{n}^B 
+ \epsilon_{H}^B(y_n) a_{n}^{B \dagger} a_{n}^B + v_n(|x_n-y_n|) a_{n}^{B \dagger} a_{n}^A \}
\end{eqnarray*} 
\end{widetext}
For the case of LUMO, we can obtain the result by $a \to b$ in $H_{el}^L$. 
Here $ a_{n}^{A \dagger} (b_{n}^{A \dagger}) $ is the electron creation operator in the HOMO (LUMO) at
the $n$th nucleotide group of P, S and $B_n$ for the chain A.
$\epsilon_{H}^A(x_n)$ and $\epsilon_{H}^B(y_n)$ 
 are the
electron-lattice coupling $H_{el-ph}$ given by
\begin{eqnarray}
\epsilon_{H}^A(x) = \epsilon_{H}^A + F_{H}x, \\
\epsilon_{H}^B(y) = \epsilon_{H}^B - F_{H}y, 
\end{eqnarray}
respectively. 
We can also obtain $\epsilon_{L}^A(x_n)$, $\epsilon_{L}^B(y_n)$ for LUMO. 
Moreover, we assume the relation:
\begin{eqnarray}
 |\epsilon_L -\epsilon_H | >> 4|t|, 
\end{eqnarray}
since this condition is realized in most of the DNA systems and guarantees the
semiconductivity of the DNA.
The total Hamiltonian $H_{tot}$ is given by
\begin{eqnarray}
H_{tot} = H_{ph} + H_{el}^H + H_{el}^L, 
\end{eqnarray}
while the wavefunction $|\Phi \rangle$ given by
\begin{eqnarray}
|\Phi \rangle = \sum_{s=A,B} \sum_{n=1}^N \{
\phi_{n}^s a^{s \dagger}_{n}|0\rangle + \phi_{n}^s b^{s \dagger}_{n}|0\rangle 
+\varphi_{n}^s a^{s \dagger}_{n}|0\rangle + \varphi_{n}^s b^{s \dagger}_{n}|0\rangle 
\}
\end{eqnarray}
where $a^{\dag}_{n\sigma}$ ($b^{\dag}_{n}$) 
means the electron creation operator at site $n$ in chain $A$ ($B$). 
 $\phi_n^A(\phi_n^B)$ represents the HOMO and $\varphi_n^A(\varphi_n^B)$ the
LUMO, at the $n$th nucleotide site in the chain A(B), respectively. 
 Applying to the Schr$\ddot{o}$dinger equation $H_{tot}|\Phi \rangle= E_{tot}|\Phi \rangle$
we obtain the following equations for the HOMO states of nucleotide groups in the
double strand of DNA:
\begin{eqnarray}
-t \{ \phi _{n+1}^A + \phi _{n-1}^A \} + \epsilon_{H}^A \phi _{n}^A +F_H x_n\phi _{n}^A -v_n \phi_{n}^B = E \phi _{n}^A\\
-t \{ \phi _{n+1}^B + \phi _{n-1}^B \} + \epsilon_{H}^B \phi _{n}^B -F_H y_n\phi _{n}^B-v_n \phi_{n}^A = E \phi _{n}^B
\end{eqnarray}
where $E \equiv E_{tot} - H_{ph}(x_n,y_n)$, $v_n \equiv v(|x_n-y_n|)$. 
For the LUMO states of nucleotide groups in the double strand of DNA, 
the similar equations are given by replacing $\phi$ and $H$ by 
$\varphi$ and $L$, respectively. 
Moreover, according to argument of Holstein, 
we find the following coupled discretized nonlinear Schr$\ddot{o}$dinger equation(DNSE)
describing the electronic behavior under the lattice vibrations in the ladder model of DNA.
In Appendix E, we give the derivation of the coupled DNSEs.
The single-chain version of the DNSE and some comments on the physical meanings
are also given in Appendix D.

\subsection{DC-Conductivity of the Double Strand of DNA}
 We exclusively focused on the low-energy transport, when the charge
injection energies are small compared with the molecular bandgap of the isolated molecule
which is of the order of $2-3eV$. 
In the experiment of the conductance property of the DNA,
temperature dependence is important.
Finite temperature can also reduce the effective system size and leads to 
the changes in the transport property.
Moreover, the effects of the stacking energy and of temperature 
can be considered by introducing the fluctuation in the hopping
energy such 
as the Su-Schriefer-Heegar model for polyacetilene \cite{su79,heeger88}. 

We consider DC-conductivity of periodic DNA sequence based on the 
LUMO, HOMO band of $\pi-$electrons and small polaron \cite{iguchi03}.
It is known that DNA behaves  $n (p)-$type extrinsic
semiconductor, where the donors (acceptors) for the LUMO (HOMO) band are
positively (negatively) charged small polarons with the total number $N_d(N_a)$ and
the energy
$\epsilon_d = E^{sp}_L = E_c -F_L^2/I_0 \omega_0^2$, 
$\epsilon_a = E^{sp}_H = E_v +F_H^2/I_0 \omega_0^2$,
where $E_v (E_c)$ is the valence (conduction) band edge.
Following the standard argument, denote by $n_c$ and $n_D$ the numbers of electrons
in the conduction (LUMO) band and in the donor levels, respectively; and
denote by $p_v$ and $p_A$ the numbers of electrons in the valence (HOMO) band and in
the acceptor levels, respectively. We now have the relation:
$n_c + n_d = N_d - N_a + p_v + p_a$, where 
\begin{eqnarray}
n_d =\frac{N_d}{ \frac{1}{2} e^{\beta(\epsilon_d-\mu)} +1}, 
p_a =\frac{N_a}{ \frac{1}{2} e^{\beta(\mu-\epsilon_d)} +1}, 
\end{eqnarray}
where $\beta\equiv 1/K_BT$, $\mu$ is chemical potential of the system.
If we suppose $\epsilon_d-\mu >>k_B T $, $\mu -\epsilon_a>>k_B T $ and 
$n_d<<N_d$, $p_a<<N_a$, then 
\begin{eqnarray}
n_c=e^{\beta(\mu-\mu_i)}n_i, 
\end{eqnarray}
where $n_i=\sqrt{N_c P_v}e^{-\beta E_g/2}$, $\mu_i=(E_c+E_v)/2 + \log (P_v/N_c)/2\beta$.

The DC conductivity  is given 
\begin{eqnarray}
\sigma = \frac{e^2 n_c \tau_e}{m_e} (=\frac{e^2p_v\tau_h}{m_h}), 
\end{eqnarray}
by Drude formula.
This suggests that if $\tau_e = const$ then the temperature dependence of $\sigma$ comes from
$n_c$, while if $n_c = const$ then it comes from $\tau_e$. 
Since the temperature dependence of $\tau_e$ comes from 
other mechanisms of scattering such as the activation of polaron
motion considered by Yoo {\it et al.}, we can assume
$\tau_e = \tau_0 e^{-\beta E_a}$
where $E_a$ is the activation energy. 
Hence, we have
\begin{eqnarray}
 \sigma = \frac{e^2 n_c\tau_0}{m_e} \exp{-\beta E_a}
\end{eqnarray}
If we adopt the $\log \sigma$ vs $1/T$ plots,
 we found the strong temperature dependence
found by Tran {\it et al.} and Yoo {\it et al.}. (See Fig.16.) 

\begin{figure}[!h]
\begin{center}
\includegraphics[scale=.5]{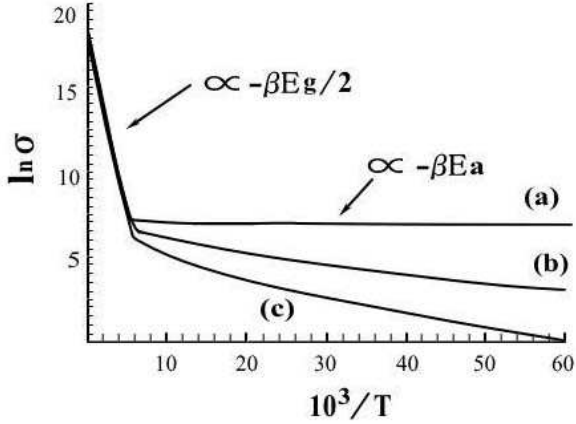}
 \caption{
The DC conductivity is shown for the n-type (or p-type) extrinsic semiconductor where
the energy gap $E_g=0.33eV$, the hopping energy for the HOMO (or LUMO) band $t=0.2eV$
and the activation energy for the polaron hopping of (a) $E_a=0.001t$, (b) $E_a=0.03t$ and
(c) $E_a=0.05t$ are taken, respectively. $N_c = P_v = 10^8$ and $n=N_D-N_A=10^3$ are assumed.
}
\end{center}
\end{figure}

\begin{figure}[!h]
\begin{center}
\includegraphics[scale=.6]{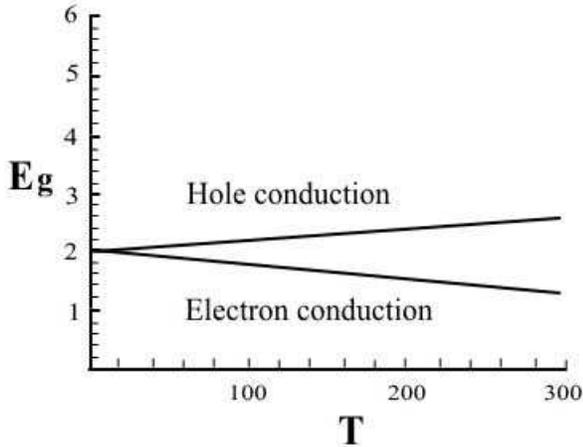}
 \caption{
Energy gap $E_g(T)$ of the $n-$type (or $p-$type) extrinsic semiconductor is shown for electron
conduction (or hole conduction) where the energy gap $E_g=2eV$, $t=0.1eV$, $N_c =P_v = 10^9$ and
$n=10^6$ are used, respectively.
}
\end{center}
\end{figure}
It is found that the band gap is
reduced by the formation of the double strand of DNA and small polarons exist to
behave as localized donors and acceptors in the DNA double helix. 
(See Fig.17.)
This extrinsic semiconducting nature of DNA qualitatively explains 
many experimental results that are observed recently.

As was stated in the introduction, 
it is very well-known that in solid state physics, 
the extrinsic semiconductor character is added by impurity atoms
that are exerted from outside into host materials.
For example, if the host material is a homogeneous crystal of $Si$ 
or $Ge$ with four valence bonds, 
then the impurities are $As$ atoms with five valence bonds
and boron(B) atoms with three valence bonds,
which then provides an inhomogeneous material with impurity levels.
The As atoms play a role of donors with positive charge such that
the system becomes an $n$-type semiconductor,
while the B atoms play a role of acceptors with negative charge
such that the system becoms a $p$-type semiconductor.
Such the system of $n$- or $p$-type semiconductor exhibits an
extrinsic semiconductor character.
On the other hand, in our systems of DNA semiconductors, 
there  exist no such impurities exerted from outside;
but already there exist inside the DNA a complicated arrangement of bases of 
A, G, C and T, 
such that among the bases, each one of bases may regard other bases as impurities.
Hence, a kind of extrinsic semiconductive nature may appear
as a result of self organization of the base arrangement.
This is the meanig of our terminology of "extrinsic semiconductor" for the DNA systems.
Therefore, it is more preferable for us to put "self-organized" in front of 
"extrinsic semiconductor" for DNA such as 
"self-organized extrinsic semiconductor" in order to represent
the semiconducting character of DNA.

We would like to note that in the $d-$dimensional disordered systems, 
the activated hopping between localized states, i.e., the variable range hopping, could 
be a dominant mechanism for the $dc-$conductivity, 
and the temperature dependence is governed by
\begin{eqnarray}
\sigma = \sigma_0 e^{-\left(\frac{T_0}{T}\right)^{\frac{d}{d+1}}}, 
\end{eqnarray}   
where $T_0=8Wa\gamma/k_B$. 
Here $W$ is the energy difference between the two sites, 
$\gamma$ is Lyapunov exponent, and $a$ the distance between the nearest neighboring bases.
It is reasonable to expect that the experiments of charge transport in $\lambda-$DNA 
suggest such temperature-dependence $\sigma \propto e^{-\sqrt{T_0/T}}$ 
at relatively low temperature due to the 1d aperiodic base-pair sequence.
The main contribution was given by the interaction with water molecules and not
with counterions. 
Further, polaron formation was not hindered by the charge-solvent coupling.
And the interaction rather increased the binding energy (self-localization) of the polaron
by around 0.5eV, which is much larger than relevant temperature scales.

Before closing this section, we give brief comments concerning the measurements 
of electronic conductivity in DNA again.
The DNA is quasi-1D systems coupled with the environment including the surrounding substrate 
and contact leads.  In general, Landauer-Buttiker formula using transfer matrix method
and Kubo-Geenwood formula using linear-response theory have been used to estimate
the transport properties in the quantum systems. 
The transfer matrix method is straightforward for the quasi-1D systems, 
however, it is inconvenient for the complicated 
cases involving the environmental effects although 
the two formulae are  equivalent at least for single particle cases \cite{datta95,cornean04}. 
There is an indication that the transfer matrix method is also effective 
even for law transmission coefficient due to the contact effect, 
despite a very good charge transfer along the DNA sequence \cite{macia06a}.

\section{Conduction and Proton Transfer}
Gene mutations sometimes cause human disorders such as cancer. 
 Simultaneously, gene mutation can be a driving force for processes of
biodiversity and evolution\cite{mcfadden01}.
When a system contains hydrogen bonds, 
proton motion also needs to be considered. 
Instead of oscillatory motions in a single-well potential, 
protons can tunnel from one side to another in a double-well
potential of the hydrogen bond. 
This proton tunneling causes an interstrand charge hopping.
And it could generate the tautomeric base pairs and
destroy the fidelity of the W-C base pairs. 

L\"{o}wdin proposed that proton tunneling contributes to the formation of rare
tautomeric of DNA W-C base pairs whose accumulation could result in
DNA damages, point mutations, and even tumor growth\cite{lowdin63}. 
The argument relies on the assumption that the rare tautomers are more stable, 
and once the intermolecular proton transfer occurs, 
lifetime of the tautomers is significantly larger than time of DNA repair. 
Indeed, the results from the quantum chemical and statistical mechanical calculations 
indicate that at room temperature at least the GC tautomers ($G^*\equiv C^*$) 
would have a sufficient lifetime to cause the DNA damage
through mismatches of the W-C base pairs.

Ladik speculated that semiconductivity of DNA might 
be related to the origin of cancer due to transmutation
of genes\cite{ladik99}. 
 Very recently, Shi {\it et al.} have also reported 
correlation between the cancereous mutation hot spots
 and the charge transport along DNA sequences\cite{shih08}. 

The proton transfer causes fluctuation of the potential and 
more or less affects on the localization/transport property of 
the charged carriers.  
At the same time, the proton transfer suffers from the lattice vibration and 
the electron transport along the base sequences of DNA.
Accordingly, we should treat the coupling between the proton transfer and  
the phononic and electronic states 
when we investigate the stability of  the tutomatic states(the excited states).
 Recent theoretical studies have shown that charged protonated base
pairs display smaller activation barriers, 
which make the proton transfer and the tunneling more facile. 

Chang {\it et al.}\cite{chang04} have given the effective 1D model Hamiltonian 
$H_{tot}=H_{el-ph}(t)+H_{\sigma}+H_{el-\sigma}$, 
where $H_{\sigma}=\sum_n H_{\sigma n}$ represents Hamiltonian describing the proton transfers. 
Here Hamiltonian $H_{el-ph}$ includes the electron and phonon modes and the coupling.
For the proton motion in hydrogen bonds, they used a
two-level system in order to describe the tunneling behavior at molecular site $n$
\begin{eqnarray}
H_{\sigma n} = \frac{1}{2}(-\epsilon_p \sigma_n^z+t_p\sigma_n^x), 
\end{eqnarray}
where 
$\sigma_n^z$ and $\sigma_n^x$ are Pauli matrices assigned at site $n$, 
$\epsilon_p$ the energy bias between the two localized proton states and 
$t_p$  the tunneling matrix element between the states.
They have modeled the coupling between the protons in the hydrogen
bond and the charges in the DNA strand as
\begin{eqnarray}
H_{el-\sigma} = \sum_n \gamma_{\sigma n} (\sigma_n^z-1) C_n^\dagger C_n 
\end{eqnarray}
where $\gamma_{\sigma n}$ is the coupling intensity. 
When the proton is in the lower energy state or there is no charge
around the hydrogen bond ($ \langle C_n^\dagger C_n \rangle = 0$), 
the coupling vanishes.

Indeed, the normal G-C base pair is in the lower energy state 
and its tautomeric form is in an excited state. 
When $\epsilon_p >>t_p$,
the probability of having a tautomeric form is extremely small. 
However, the cation of a G-C base pair has almost the same energy 
as that of its tautomeric form $ \epsilon_p \leq t_p$. 
In this case, the proton state becomes delocalized in hydrogen bonds.

The symmetry of potential well and the hight of barrier essentially affects on the 
tunneling probability between the potential wells.
The probabilistic amplitude of proton could influence the transport of the
charged carriers though the Coulomb interaction.
In addition, the two-level approximation will be broken down
 if the chaotic motion occurs in the dynamics due to 
the coupling with lattice vibrations \cite{igarashi08}.


\section{Summary and Discussion}
We briefly reviewed our recent works with concerning the electronic states and 
the conduction/transfer in DNA polymers.  
In Sect.3,  based on the H$\ddot{u}$ckel model,
we have discussed the basics of quantum chemistry 
where the electronic states of atoms in biochemistry such as C, N, O, P 
and the electronic configurations of the nitrogenous bases, the sugar-phosphate groups, 
the nucleotides and the nucleotide bases are summarized, respectively. 
In Sect.4, based on the tight-binding approximation of the DNA sequences,  
we have investigated the localization properties of electronic states and quantum diffusion,  
using a stochastic-bond-vibration approach for the poly(dG)-poly(dC),
the poly(dA)-poly(dT) and the mixed DNA polymers within the frameworks of the polaron models. 
At that time, we assumed that the dynamical disorder is caused by the DNA vibrational modes, 
which are caused by a noise associated with stochastic dynamics 
of the distances $r(t)$ between the two Watson-Crick base pair partners.
In Sect.5, 
we have mainly investigated the energy band structure and 
the localization property of electrons in the disordered 
two- and three-chain systems (ladder model) with
long-range correlation as a simple model for electronic property 
in a double strand of DNA by using the tight-binding model. 
In addition, we investigated the
localization properties of electronic states in several actual 
DNA sequences such as bacteriophages of Escherichia coli, human
chromosome 22, compared with those of the artificial disordered sequences. 
In Sect.2 and 4, we gave a brief explanation for DNA bases sequences and gene mutations,
respectively, 
which are related to the DNA carrier conduction phenomena
\cite{gutierrez05,gutierrez06,macia06,macia07}. 
 The relationship between the long-range correlations and the coherent charge transfer 
in the substitutional DNA sequences has been studied, 
based on the transfer matrix approaches\cite{guo07,bagci07}.
 
 Recently, the tight-binding models for the DNA polymers have been 
extended to the decorated ladder models and the damaged DNA models by some groups.
Furthermore, the I-V characteristics of the ladder models coupled with environments
has been investigated as simple models for DNA polymers \cite{malyshev07,ketabi09,gutierrez09}.



\appendix

\section{H\"{u}ckel Parameters}
Sandorfy's formula\cite{sandorfy49}, Streitwieser's formula\cite{streitwieser61},
and Mulliken's formula\cite{mulliken49} are given by following relations,
\begin{eqnarray}
\alpha_{\dot{X}} &=& \alpha + \frac{\chi_{X}-\chi_{C}} 
{\chi_{C}}\times 4.1\beta, \\
 \alpha_{\ddot{X}} &=& \alpha_{\dot{X}}  + \beta, \\
 \ell_{CX} &= & \frac{S_{CX}}{S_{CC}},
\end{eqnarray}
respectively.
See Ref.\cite{iguchi04} for the details.

\section{Transfer Matrix Method}
Let us define the four-dimensional column vector  in the main text as
$\Phi_{n} \equiv (\phi_{n}^A, \phi_{n-1}^A, \phi_{n}^B, \phi_{n-1}^B)^{t}$.  
Then, it can be rewritten in the following form:
$\Phi_{n+1}  = M_{n} \Phi_{n}$, 
where
\begin{widetext}
\begin{eqnarray} 
M_{n}   =  
\left( \begin{array}{cc}
A_{n}&V_{n}\\  
U_{n}&B_{n}
\end{array}
\right).
\end{eqnarray} 
$M_{n}$ is the $4\times4$ transfer matrix with the $2\times2$ matrices:
\begin{eqnarray}
A_{n}    & \equiv &  
\left( \begin{array}{cc}
\frac{E - A_{n,n}}{A_{n+1,n}}&-\frac{A_{n,n-1}}{A_{n+1,n}}\\
1&0
\end{array}
\right), 
 B_{n}   \equiv   
\left( \begin{array}{cc}
\frac{E - B_{n,n}}{B_{n+1,n}} &-\frac{B_{n,n-1}}{B_{n+1,n}}\\
1&0
\end{array}
\right),\\  
U_{n}    & \equiv &   
\left( \begin{array}{cc}
\frac{-U_{n}}{A_{n+1,n}}&0\\
0&0
\end{array}
\right),   
 V_{n}    \equiv    
\left( \begin{array}{cc}
\frac{-V_{n}}{B_{n+1,n}}&0\\
0&0
\end{array}
\right).                                                                 
\end{eqnarray}
\end{widetext}
According to the sequence of $N$ segments, we have to take the matrix 
product $M(N)   \equiv  M_{N} M_{N-1} \cdots M_{1}$,
which is also a $4\times4$ matrix.  
When the double chain system is periodic,
we adopt the Bloch theorem:
\begin{eqnarray}
\phi_{n+N}^A  = \rho\phi_{n}^A,\ \ \ \phi_{n+N}^B = \rho \phi_{n}^B,    
\end{eqnarray}
where $\rho = e^{ikN}$ and $k$ is the wavevector. 
Then, we find a $4\times4$  determinant $D(\rho)$ that is a fourth order polynomial of $\rho$:  
\begin{eqnarray}
D(\rho) \equiv  \det[M(N) - \rho I_{4}] \nonumber \\
= \rho^{4}  - a_{1}\rho^{3} + a_{2}\rho^{2}  - a_{3}\rho + a_{4}  = 0.                           
\end{eqnarray}
It provides the wave vector $k$ in the system such that $k = 2\pi j/N$ 
for $j = -N/2, \cdots , N/2$.   
Here, if four roots are written as $\rho_{1}$, $\rho_{2}$, $\rho_{3}$, 
and $\rho_{4}$, then the following conditions hold 
\begin{eqnarray}
a_{1}  \equiv  tr M(N)  = \sum_{i=1}^{4}\rho_{i}, 
a_{2}  \equiv  \sum_{i<j=1}^{4} \rho_{i} \rho_{j}, \nonumber \\
a_{3}  \equiv  \sum_{i<j<k=1}^{4}\rho_{i}\rho_{j} \rho_{k}, 
a_{4}  \equiv  \det M(N)  = \rho_{1}\rho_{2}\rho_{3}\rho_{4}. 
\label{a1234}
\end{eqnarray}

Using a physical intuition, if an electron propagation with $k$ along one direction 
in the double chain is represented by $\rho$,
then the reverse propagation with $-k$ is represented by $\rho^{-1}$.  
Therefore, the latter should be also accessible,
since the choice of the direction of the coordinate system is arbitrary.  
Hence, $\rho^{-1}$ must be an eigenvalue of $D(\rho) = 0$ such that 
\begin{eqnarray}
D(\rho^{-1}) = \rho^{-4}(a_{4}\rho^{4} - a_{3}\rho^{3} 
+ a_{2}\rho^{2} - a_{1}\rho + 1)  = 0.                        
\end{eqnarray}
This situation imposes the particular condition on the matrix $M$:   
\begin{eqnarray}
M^{\dag} J M  =  J,                                     
\end{eqnarray}

\begin{eqnarray}
J  \equiv  
\left( \begin{array}{cc}
\textbf{J}&\textbf{0}\\
\textbf{0}&\textbf{J}
\end{array}
\right),\ \   
\textbf{J}  \equiv 
\left( \begin{array}{cc}
0&-1\\
1&0
\end{array}
\right),
\end{eqnarray}                                                        
where $M^{\dag}$ means the Hermitian conjugate of $M$ and $\textbf{0}$ is the 
$2\times2$ zero matrix, respectively.   
This property is called the symplectic structure of $M$, 
where we have 
\begin{eqnarray}
D(\rho)  = \rho^{4}D(\rho^{-1}),                               
\end{eqnarray}
from which we find 
$a_{1}  = a_{3},\ \ a_{4} = 1$.                                
Thus, $M$ belongs to $SL(4, \textbf{R})$.
By using this property and dividing $D(\rho)$ by $\rho^{2}$, 
the biquadratic equation is reduced to the quadratic equation: 
\begin{eqnarray}
x^{2}  - a_{1}x + a_{2} - 2 = 0,    x  =  \rho  +  \frac{1}{\rho} 
\end{eqnarray}
Therefore, its two roots are given as 
\begin{eqnarray}
x_{\pm}  =   \frac{1}{2}  \left(a_{1} \pm \sqrt{D}\right),\ \   
D = a_{1}^{2} - 4a_{2} + 8.                                        
\end{eqnarray}
Since from Eq.(\ref{a1234}) and
$tr(M^{2}) = \sum_{i=1}^{4}\rho_{i}^{2}$, 
we obtain 
$a_{1}^{2} - 2a_{2} = tr(M^{2})$.   

Now we can state a simple scheme to obtain the spectrum: 
If an energy $E$ satisfies 
\begin{eqnarray}
x_{\pm} = 2 \cos kN, 
\end{eqnarray}                                              
then the energy is allowed, otherwise it is forbidden in channel $\pm$, respectively.  
This is a generalized version of the Bloch condition for the single linear 
chain system with the $2\times2$ transfer
matrix $M$ where 
\begin{eqnarray}
tr M = 2 \cos kN.                                                    
\end{eqnarray}
The density of states (DOS) $D_{\pm}(E)$ is calculated 
 for each channel $\pm$, respectively:
\begin{eqnarray} 
dk_{\pm}  &=&  \frac{1}{N}   d \cos^{-1}[x_{\pm}(E)/2] \\
  &=&  - \frac{1}{N}  \frac{\frac{\partial x_{\pm}}{\partial \varepsilon}}
{\sqrt{4 - x_{\pm}^{2}}}  dE  =  D_{\pm}(E) dE. 
\end{eqnarray}
Therefore, the total DOS is given as the sum of $D_{+}(E)$ and $D_{-}(E)$:
\begin{eqnarray}
D(E) = D_{+}(E) + D_{-}(E),        
\end{eqnarray}
where $D_{-}(E)$ [$D_{+}(E)$] means the DOS contributed from 
the bonding (antibonding) channel $-$ ($+$), respectively.
It agrees with the result on the tight-binding model for the ladder structure.  
Physically speaking, the $-$ ($+$) channel means the bonding (antibonding) states
between two parallel strands of the DNA.

\section{Transfer Matrices for Ladder Systems}
 In this appendix, we give the explicit expression of the energy band for
 the decolated ladder model given in  Fig.10(a).
 In the unit cell, the period is taken as $N=2$ and it contains four $\pi-$orbitals.
 We apply the result in Appendix B for the case. 
Let the transfer matrix method be 
\begin{eqnarray}
M   =  
\left(
\begin{array}{cc}
A&V\\  
U&B
\end{array}
\right).                                                   
\end{eqnarray}
 $M$ is a $4\times4$ transfer matrix with $2\times 2$ matrices:
 
\begin{eqnarray}
A &=& A_{n+1}A_{n}   
\equiv   
\left(\begin{array}{cc}
\frac{(E - \beta)(E - \alpha)}{ab} -\frac{a}{b} & -\frac{E-\beta}{a} \\
 \frac{E-\alpha}{a} & -\frac{b}{a}   
\end{array}
\right) =B, \\  
V &=& A_{n+1}V_{n}     
\equiv   
\left(
\begin{array}{cc}
- \frac{E-\beta}{ab}v & 0 \\
-\frac{v}{a} & 0
\end{array}
\right) =U. 
\end{eqnarray}

Let us calculate $trM$ and $tr(M^{2})$.
We find
\begin{eqnarray}
trM &=& tr(A) + tr(B) = 2P -2R \\
tr(M^{2}) &=& tr(A^2) + tr(B^2) + tr(UV) + tr(VU) \nonumber \\
&=& 2(P+Q)^2 +2(R^2-2), 
\end{eqnarray}
where 
\begin{eqnarray}
P = \frac{(E - \beta)(E - \alpha)}{ab}, 
Q = \frac{(E - \beta) v}{ab} , 
R = \frac{a^2+b^2}{ab}.
\end{eqnarray}
The discriminant $D$ is given by $D = 2 tr(M^{2}) - (tr M)^{2} + 8 = 4Q^2$. 
As a result, we obtain
\begin{eqnarray}
2\cos 2k &=& \frac{1}{2} (trM \pm \sqrt{D}) \\
 &=& P-R \pm Q \\
 &=& \frac{(E - \beta)(E - \alpha)}{ab}- \frac{a^2+b^2}{ab}   \pm  \frac{(E - \beta) v}{ab}
\end{eqnarray}
Solving the above for $E$, we can obtain the energy bands.
\begin{widetext}
\begin{eqnarray}
 E_{+}^{(\pm)} &=& \frac{1}{2} \left( (\alpha+\beta) - v \pm 
\sqrt{ (\alpha-\beta)^2+v^2+4(a^2+b^2) + 2(\alpha+\beta) v +4\beta v
+ 8ab\cos 2k }   \right)  \\
 E_{-}^{(\pm)} &=& \frac{1}{2} \left( (\alpha+\beta) + v + \pm
\sqrt{ (\alpha-\beta)^2+v^2+4(a^2+b^2) - 2(\alpha+\beta) v -4\beta v
+ 8ab\cos 2k }   \right)  
\end{eqnarray}
\end{widetext}
In Fig.1(a), the energy band structure for $a=b=1$ when $\alpha=1, \beta=0$
is given with varying the interchain hopping $v$.
Figure 1(b) shows the cross-section view at $v=1$.

\begin{figure}[!h]
\begin{center}
\includegraphics[scale=.6]{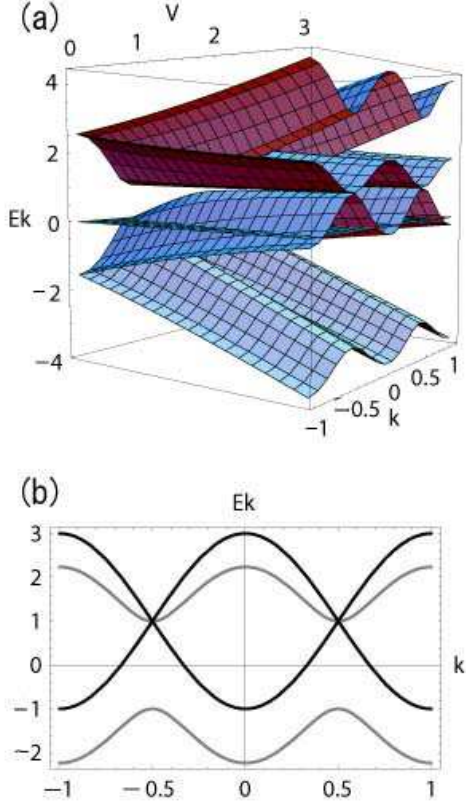}
\caption{
Energy bands of the decorated ladder model.
(a) The energy bands as a function of $V$.
(b) The snap shot of the energy bands when $V = 1$.
$k$ means the wave vector in units of $\frac{\pi}{s}$ such that $-1.0 \le k \le 1.0$,
$V$ means the $\pi$-electron hopping integral between
the inter chain sites, and
$E$ means the energy in units of $V = 1$.
Here we have taken the values 
$\alpha=1.0$, $\beta= 0$, $a=b=1$.
}
\end{center}
\end{figure}

The extension to the decolated three chains ($d=3$) is straightforward.
Here we give the transfer matrix for a simple system of coupled three chains.
$\Phi_{n+1}= M_n \Phi_n$
where 
$\Phi_{n}=(\phi^{A}_{n+1},\phi^{A}_{n},\phi^{C}_{n+1},\phi^{C}_{n},\phi^{B}_{n+1},\phi^{B}_{n})^t$.
$6\times6$ transfer matrix with the $2\times 2$ submatrices is given in block tridiagonal form: 
\begin{eqnarray}
M_n   =  
\left( 
\begin{array}{ccc}
A_n &V_n^A & 0\\  
V_n^C & C_n & U_n^C \\
0 & U_n^B & B_n 
\end{array}
\right),                                                    
\end{eqnarray}
 where 
\begin{widetext} 
\begin{eqnarray*}
C_{n}    \equiv   
\left( \begin{array}{cc}
\frac{E - C_{n,n}}{C_{n+1,n}} & -\frac{C_{n,n-1}}{C_{n+1,n}}\\
1&0
\end{array}
\right),  
 V_{n}^A   \equiv   
\left( \begin{array}{cc}
-\frac{V_{n}}{A_{n+1,n}} &0 \\
0&0
\end{array}
\right), 
 V_{n}^C    \equiv    
\left( \begin{array}{cc}
-\frac{V_{n}}{C_{n+1,n}} &0 \\
0&0
\end{array}
\right), \\ 
U_{n}^C     \equiv    
\left( \begin{array}{cc}
-\frac{U_{n}}{C_{n+1,n}}&0\\
0&0
\end{array}
\right),   
 U_{n}^B    \equiv    
\left( \begin{array}{cc}
-\frac{U_{n}}{B_{n+1,n}}&0\\
0&0
\end{array}
\right).                                                                 
\end{eqnarray*}
\end{widetext}
For simplicity, we set $A_{n+1,n} = B_{n+1,n}=a$, $C_{n+1,n}=c$,   
 $A_{nn}=B_{nn}=\alpha$, $C_{nn}=\kappa$,
and $V_n=U_n=v$ at site $n$. Then 
\begin{eqnarray}
M   =  
\left( \begin{array}{cccccc}
\frac{E-\alpha}{a} & -1 & -\frac{v}{a} & 0 & 0 & 0 \\  
1 & 0 & 0 &0 &0 &0 \\
-\frac{v}{c}& 0 & \frac{E-\kappa}{c} & -1 & -\frac{v}{c} & 0 \\  
 0 & 0& 1 & 0& 0& 0 \\
0&0& -\frac{v}{a} & 0 & \frac{E-\alpha}{a} & -1  \\
0& 0& 0& 0&  1&0 
\end{array}
\right).                                                    
\end{eqnarray}

Even for general multichain models,
some useful formula exist in order 
to obtain the eigenvalues of block tridiagonal matrices and 
the determinants of the corresponding block tridiagonal matrices 
\cite{hori68,schenk06,molinari08}.

\section{Nonlinear Schr$\ddot{o}$dinger Equation} 
In this appendix, we derive the discrete nonlinear Schr$\ddot{o}$dinger equation
for 1D electronic system coupled with lattice oscillations.
The relative motions between two
different base pairs can be represented by an acoustical
phonon mode and the vibrational motion inside a base
pair by optical phonons. 
These modes represent the lattice distortions
such as sliding, twisting or bending. 
The Hamiltonian that describes these modes is given by
\begin{widetext}
\begin{eqnarray}
H_{ph} = \sum_n\{ \frac{p_n^2}{2M} +\frac{1}{2}M\omega_s^2(u_{n+1}-u_n)^2  \} 
 + \sum_n\{ \frac{P_n^2}{2M} +\frac{1}{2}M\omega_o^2v_n^2  \}.
\end{eqnarray}
\end{widetext}
Here $u_n$ and $v_n$ are lattice displacements and the internal
vibration coordinates of the $n$th unit cell
and $p_n$ and $P_n$
are their conjugated momentum, respectively.
And $M$ is the mass of the unit cell, 
$\omega_o$ is the oscillation frequency of the optical phonon 
and the dispersion relation of the acoustical motion 
is $\omega_s (k) = c_s k$, where $c_s$ is the sound velocity
along the chain.
 Note that the two kinds of oscillations can be regarded as the dynamics of 
 radial and angular coordinates in the polaron models in Sect.4.

Then in the Hamiltonian for electrons, 
both the on-site potentials $E_n$ and the hopping integrals $V_{m,n}$  
depend upon these vibrations,  in principle. 
 The charge coupling to the acoustical phonons is given
by the Su-Schrieffer-Heeger (SSH) model \cite{su79,heeger88} and the interaction
with the optical phonons is described via the molecular crystal
model of Holstein \cite{holstein59}. 
The SSH model deals classically with the lattice degrees of freedom,
while electrons are treated quantum mechanically.
Thus the total Hamilonian for the electron-phonon
interactions in the DNA system is given by
\begin{widetext}
\begin{eqnarray}
  H_{el-ph} &=& \sum_n \alpha (V_0+ u_{n+1}-u_n+\delta V_{nn+1}) 
(C_n^\dagger C_{n+1} + C_{n+1}^\dagger C_n) 
+ \sum_n (E_n+\gamma_n v_n) C_n^\dagger C_n, 
\end{eqnarray}
\end{widetext}
where $\alpha$ and $\gamma_n$ are the coupling constants. 
 Here $V_0$ is the bare amplitude of the hopping term and 
the term $\delta V_{nn+1}$ is a random contribution
from the conformational disorder, which we include to describe.
 Anderson localization takes place when $E_n$ is a static on-site randomness.  
 When $E_n=constant$  in the system
with half-filled conduction band, such as an 1D ionic crystal, 
 the SSH term generates
dimerization in the ground state (the Peierls instability)
and forms solitons in the excited states. 

However, DNA is considered as a band insulator.
Both interactions can generate lattice distortions and lead to polaron
formation when a charge is doped into the molecule. 

The coupling with $(u_{n+1}-u_n)$ usually induces small polarons in ionic crystals.
 The calculations showed that a polaron may be built and be robust within 
a wide range of model parameters. 
The influence of random base sequences was apparently not strong enough to destroy it. 
Thus, polaron drifting may constitute a possible transport mechanism in DNA oligomers.

From the total Hamiltonian $H_{tot}=H_{ph}+H_{el-ph}$ of the system
we can derive the equations of motion for the variables $\phi_n$, $u_n$ and $v_n$
as follows:
\begin{widetext}
\begin{eqnarray}  
\frac{d^2 u_n}{dt^2} &+& \omega_s^2(2u_n(t)-u_{n+1}(t)-u_{n-1}(t))= 
2\alpha Re(\phi_{n+1}^{*}(t)\phi_{n}(t) \phi_{n}~{*}(t)\phi_{n-1}(t)) \\
\frac{d^2 v_n}{dt^2} &+& \omega_o^2v_n(t) =  \gamma_n|\phi_n(t)|^2  \label{opt} \\
i\hbar \frac{d \phi_n}{dt} &=&  (E_n+\gamma_n v_n) \phi_n(t) + \sum_m t_{mn} \phi_m(t),
\end{eqnarray} 
\end{widetext} 
where $t_{mn}=(V_0+ u_{m}-u_n+\delta V_{mn})$. 
This means that the equilibrium position of each atom 
in the lattice is charged by an amount proportional to the probability 
for the electron to occupy that special atom. 
   The twist polaron modify the inter-base electronic coupling, 
though this effect is apparently less strong than the coupling in the Holstein model.
 Accordingly, we also assume that the dependence of the hopping integrals on the $u_n$ is so week
so that it can be ignored.
In the following part, we deal with only the Holstein polarons.

Additionally, we assume that time scale of lattice vibrations and electron evolution
are as such that vibrations are slaved by electron probability.
It is then possible to set time derivative in Eq.(\ref{opt}) equal to zero.
Then, we obtain
\begin{eqnarray}  
i\hbar \frac{d \phi_n}{dt} =   \sum_m t_{mn} \phi_m(t) +  
\alpha |\phi_n(t)|^2 \phi_n(t) +E_n \phi_n(t),
\end{eqnarray}  
where $\alpha=\gamma_n^2/\omega_o^2$.
By setting $t_{mn}=-t$ if $|m-n|=1$ and zero otherwise, 
we obtain the discrete nonlinear Schr$\ddot{o}$dinger equation (DNSE), 
\begin{widetext}
\begin{eqnarray}  
i\hbar \frac{d \phi_n}{dt} =   - t(\phi_{n+1}(t) + \phi_{n-1}(t) ) +  
a|\phi_n(t)|^2 \phi_n(t) +E_n \phi_n(t) .
\label{DNSE1}
\end{eqnarray}  
\end{widetext}
The time-independent version is given as
\begin{eqnarray}  
E\phi_n = - t(\phi_{n+1}+\phi_{n-1}) +\alpha |\phi_n|^2\phi_n +E_n\phi_n, 
\label{DNSE2}
\end{eqnarray},  
which appears in the Holstein polaron model on the lattice.
The DNSEs (40) and (41) show more various properties
due to the inhomogeneity 
of $E_n$ and the strength $\alpha$ for nonlinearity, etc \cite{kopidakis08,flach09}.

For example, localization-delocalization transition takes place, depending on
the coupling strength and the initial state.
The DNSE has been studied in the context of delocalization due to nonlinearity. 
Indeed, Eq.(\ref{DNSE1}) describes the 1D disordered waveguide lattice, 
which is called the Gross-Pitaevsky (GP) equation on a discretized lattice \cite{pikovsky08}.
When the absolute value of the nonlinearity parameter $a$ is greater than some
critical value $\alpha_c$, the excitation is self-trapped.
It was mathematically proved that the DNLS has quasi-periodic self-trapped solutions
called the discrete breathers\cite{lahini08,larcher09}.
 On the other hand, it is found that at moderate strength of nonlinearity 
the spreading of the wavepacket algebraically grows as 
$\surd \langle \Delta n)^2 \rangle \sim t^\nu (\mu \sim 0.2)$ \cite{garcia09}.

\section{Coupled Nonlinear Schr$\ddot{o}$dinger Equations}
In this appendix, we derive the coupled nonlinear Schr$\ddot{o}$dinger equations
in order to describe the effect of  formation of a double strand of DNA on 
the polarons in the HOMO band, following the Holstein's argument\cite{iguchi03}.
The energy of HOMO band is expressed as, 
\begin{widetext}
\begin{eqnarray}  
E_H(\{ x_n, y_n\}) = H_{ph}(\{ x_n, y_n\}) + 
\sum_{n=1}^N \left( \epsilon_H(x_n) |\phi_n^A|^2 + \epsilon_H(y_n) |\phi_n^B|^2 \right)  \nonumber \\
-  \sum_{n=1}^N t\left((\phi_{n+1}^A+ \phi_{n-1}^A)\phi_{n}^{B*} +
  (\phi_{n+1}^B+ \phi_{n-1}^B)\phi_{n}^{A*} \right) 
-\sum_{n=1}^N v_n(\phi_n^{A} \phi_n^{B*} +\phi_n^{B} \phi_n^{A*}).
\end{eqnarray}
\end{widetext}
Differentiating the energy with respect to $x_p$, $y_p$, respectively, 
we approximately obtain  the most contributed coordinates $x_p, y_p$ by 
$\frac{\partial E_H}{\partial x_p} =0$, $\frac{\partial E_H}{\partial y_p} =0$;
\begin{eqnarray}  
x_p &=& -\rho |\phi_p^A |^2 -\nu (\phi_{p}^A \phi_{p}^{B*} + \phi_{p}^B \phi_{p}^{A*}), \\
y_p &=& -\rho |\phi_p^B |^2 -\nu (\phi_{p}^A \phi_{p}^{B*} + \phi_{p}^B \phi_{p}^{A*}), 
\end{eqnarray} 
 where $\rho=F_H/I_0\omega_0^2$, $\nu=\alpha/I_0\omega_0^2$. 
In the derivation,  we have assumed $v_p=v-\alpha|x_p-y_p|$,  $\frac{\partial S_H^{AB}}{\partial x_p} =0$,
$\frac{\partial S_H^{AB}}{\partial y_p} =0$, where 
$S_H^{AB}=(\phi_{p}^A \phi_{p}^{B*} + \phi_{p}^B \phi_{p}^{A*})$.
Note that $|\phi_n^A|^2 $($|\phi_n^B|^2$) means the frontier orbital density 
at $p$-th nucleotide group in chain A(B)
and $S_H^{AB}$ means the overlapping integral of the frontier orbitals for electrons in the HOMO at 
$p$-th nucleotide group between the chains A and B.
Substituting the expressions into Eq.(34) and (35), 
we obtain the following  coupled DNSEs:
\begin{eqnarray}
-t \{ \phi _{n+1}^A + \phi _{n-1}^A \} + \epsilon_{H}^{AB} \phi _{n}^A  -v_n^{AB} \phi_{n}^B = E \phi _{n}^A\\
-t \{ \phi _{n+1}^B + \phi _{n-1}^B \} + \epsilon_{H}^{BA} \phi _{n}^B -v_n^{AB} \phi_{n}^A = E \phi _{n}^B
\end{eqnarray}
where 
\begin{eqnarray}  
\epsilon_{H n}^{AB} &=& \epsilon_{H}^A -\rho F_H |\phi_n^A |^2 +\nu F_H S_H^{AB}, \\
\epsilon_{H n}^{BA} &=& \epsilon_{H}^B -\rho F_H |\phi_n^B |^2 -\nu F_H S_H^{AB}, \\
v_{n}^{AB} &=& v -\nu F_H (|\phi_n^A |^2 +|\phi_n^B |^2). 
\end{eqnarray} 
 By the same argument, similar equations can be obtained for the LUMO band case as well, 
 just by replacing the orbitals of electrons in HOMO bands  with the  
 frontier orbitals for holes in the LUMO bands, respectively.
In the limit $v \to 0, \nu \to 0$, it becomes the decoupled 
DNSE in Eq.(\ref{DNSE2}) without randomness.

\section{Lyapunov Exponents and Multichannel Conductance}
 The definition for  energy dependence of the Lyapunov exponents
 is given by,
\begin{eqnarray}  
\gamma_i = \lim_{N \to \infty} \frac{1}{2N} \log \sigma_i(M_d(N)^\dagger M_d(N)),    
\end{eqnarray}  
\noindent
where $\sigma_i(\dots) $ denotes the $i$-th eigenvalue of the matrix $M_d(N)^\dagger M_d(N)$  \cite{yamada01}. 
 As the transfer matrix $T_d(N)$ is symplectic, the eigenvalues of the
$M_d(N)^\dagger M_d(N)$ have reciprocal symmetry around the unity as
$e^{\gamma_1},...,e^{\gamma_d} e^{-\gamma_d},...,e^{-\gamma_1}$, 
where $\gamma_1 \geq \gamma_2\geq \dots \gamma_d \geq  0$.
 The $d$ denotes the number of channels;
i.e. $d=2$ in the two-chain model
and $d=3$ in the three-chain model.

The Lyapunov exponent is related to the DOS $\rho(E)$ as an analogue of the Thouless
relation (the generalized Thouless relation)  \cite{molinari03}:
\begin{eqnarray}
 \sum_i^d \gamma_i(E)  \sim \int \ln |E-E^{'}| \rho(E^{'}) dE^{'}.   
\end{eqnarray}
\noindent
Accordingly, we can see that 
the singularity in the largest Lyapunov exponent is strongly related to the 
singularity in the DOS.

Generally speaking, in the quasi-one-dimensional chain 
with the hopping disorder, 
the singularity in the DOS, the localization length and the 
conductance at the band center depend on the parity, the bipartiteness 
and the boundary condition. 
Since discussions on the details is out of scope of this paper, 
we give simple comments.    
Note that the parity effects appear in the odd number chains 
with the hopping randomness.
In the odd number chain with the hopping randomness, 
only one mode at $E=0$  is remained as an extended state, i.e. $\gamma_d=0 $, 
while other exponents are positive, $\gamma_{d-1} > \dots > \gamma_1 >0$.
 The behavior is seen in Fig.14 in the main text.
 Then,  non-localized states with $\gamma =0$ determine the conductance.
  Although we have ignored the bipartite structure in the three-chain models for simplicity, 
if we introduce bipartiteness in the intrachain hopping integral $V_n(=U_n)$ 
another delocalized state due to chiral symmetry appears at $E=0$.

Furthermore, we find that in the thermodynamic limit ($n \to \infty$),
the largest channel-dependent localization length $\xi_d=1/\gamma_d$ 
determines the exponential decay in the Landauer conductance $g(n)$
that is measured in units of $e^2/h$ at zero temperature and serves as the localization
length of the total system of the coupled chains \cite{lifshits88,imry02}.
It is given as
\begin{eqnarray}
 g(n) = 2 \sum_{i=1}^{d} \frac{1}{\cosh \frac{2n}{\xi_i(n)} -1} \sim \exp (-\frac{2n}{\xi_d(n)} ),
\end{eqnarray}
\noindent
for $n \to \infty $, 
where $n$ denotes the system size along the chain \cite{crisanti93}. 
Recently, electron transport for molecular wires between two metallic electrorodes
has been also investigated by several techniques. 
\\
If we impose a voltage difference $V$ over the molecular structure,
we have one end point with energy $E_F+eV$ and the other still $E_F$.
Only electrons in that range contribute to the conductance.
Therefore, the $I-V$ characteristics can be evaluated as
\begin{eqnarray}
 I(E_F) = \frac{Ve}{\pi \hbar} \int dE (-\frac{\partial f}{\partial E}) \sum_i T_i(E) \\
 \to  \frac{Ve}{\pi \hbar} \sum_{ij} T_{ij}(E_F), 
\end{eqnarray}
\noindent 
where $f(E)\equiv 1/(e^{E/k_BT}+1)$ means the Fermi distribution function and 
$T_{ij} \equiv |t_{ij}|^2$ is the squared transmission amplitudes between the $i$-th and $j$-th channels. 
At the zero-temperature limit,  all quantities can be evaluated by the transmittion matrix at $E=E_F$.

\section{Modified Bernoulli Map}
The correlated binary sequence $\{V_n\}$ and/or $\{C_{nn+1} \}$ 
of the hopping integrals can be generated by 
 the modified Bernoulli map 
\cite{aizawa84,yamada91}:

\begin{eqnarray}
 X_{n+1}= 
\left\{ \begin{array}{ll}   
   X_{n} + 2^{B_0-1}X_{n}^{B_0}  & (X_{n} \in I_0)  \\
   X_{n} - 2^{B_1-1}(1-X_{n})^{B_1} & (X_{n} \in I_1), \\   
\end{array} \right.   
\label{eq:map}
\end{eqnarray}
\noindent
where $I_0=[0,1/2),I_1=[1/2,1)$.  
$B_0$ and $B_1$ are the bifurcation parameters that control correlation in the sequence. 
We set $1<B_0<B_1<2$ for simplicity. 
The asymmetry of the map ($B_0 \neq B_1$) corresponds to the asymmetric property 
in the distribution for a real sequence in the double helix DNA,
where the number of the A-T pairs is not equal  to that of G-C pairs; 
they are different from random binary sequences with equal weight. 
We introduce an indicator $R_{GC}$ for the rate of the G-C pairs in the sequences such as
$R_{GC} = (N_G+N_C)/(N_G+N_C +N_A+N_T)$, 
where $N_G, N_C, N_A$ and $N_T$ denote the numbers
of symbols G, C,  A and T in the base sequence, respectively.

In the case of  $B_0=B_1(\equiv B)$, depending on the value $B$,
the correlation function $C(n)(\equiv \langle V_{n_0}V_{n_0 + n}\rangle)$ ($n_0=1$, $n$ is even ) 
decays following inverse power-law as 
$C(n)\sim n^{-\frac{2-B}{B-1}}$ for large $n$ ($3/2<B<2$). 
The power spectrum becomes 
$S(f)\sim f^{-\frac{2B-3}{B-1}}$ for small $f$. 
We focus on the Gaussian and non-Gaussian stationary regions ($1<B<2$) 
that correspond to some real DNA base-pair sequences with 
$S(f) \sim f^{-\alpha} (0.2<\alpha<0.8)$.  

In the ladder models of Sect. 5 , 
we use  for the interchain hopping integrals at odd sites $n$
the symbolized sequence $\{ V_{n} \}$  that are defined by the following rule:
\begin{eqnarray}
V_{n} =  
\left\{ \begin{array}{ll}
W_{AT} = W_{TA} & (X_{n} \in I_0) \\
W_{GC} = W_{CG} & (X_{n} \in I_1). \\
\end{array} \right.
\end{eqnarray} 
\noindent
(See Fig.10(a).)
In the numerical calculation, $W_{GC}$ is set at a half of $W_{AT}$
($W_{GC}=W_{AT}/2$) for simplicity.
Then, the artificial binary sequence can be roughly regarded as the base-pair
sequence as observed in the $\lambda-$DNA. 
 Note that the correlated four-letter sequence $\{ E_n \}$ such as DNA 
can be also generated by using two independent sequences $\{ X_n \}$, $\{ Y_n \}$ 
 by the modified Bernoulli map:
 \begin{eqnarray}
E_{n} =  
\left\{ \begin{array}{ll}
A & (X_{n} \in I_0, Y_{n} \in I_0) \\
T & (X_{n} \in I_0, Y_{n} \in I_1) \\
G & (X_{n} \in I_1, Y_{n} \in I_0) \\
C & (X_{n} \in I_1, Y_{n} \in I_1). \\
\end{array} \right.
\end{eqnarray}

 In the three-chain models, 
 the interchain hopping integral $V_n=U_n$ for every site $n$ can be generated in the same way
tas to the two-chain models.
 Furthermore, we use the successive sequence $\{X_n,X_{n+1} \}$,
when we make a correlated binary sequence 
$\{ C_{nn+1} \}$ as the hopping integral of 
the middle (nucleotide) chain as follows:
\begin{eqnarray}
C_{nn+1} =  
\left\{ \begin{array}{ll}
W_{AT-AT} & (X_{n} \in I_0, X_{n+1} \in I_0) \\
W_{AT-GC} & (X_{n} \in I_0, X_{n+1} \in I_1) \\
W_{GC-AT} & (X_{n} \in I_1, X_{n+1} \in I_0) \\
W_{GC-GC} & (X_{n} \in I_1, X_{n+1} \in I_1). \\
\end{array} \right.
\end{eqnarray} 
\noindent
In the numerical calculation, we assume the following rules for simplicity:
\begin{eqnarray}
\left\{ \begin{array}{l}
W_{GC-GC} = W_{AT-AT}/2 \\
W_{AT-GC} = (W_{AT-AT} + W_{GC-GC})/2 \\
W_{GC-AT} =  W_{AT-GC}. \\
\end{array} \right.
\end{eqnarray} 
\noindent
As a result, the parameters are $W_{AT}$ and $W_{AT-AT}$.
This simple rule is based on the binary classification of the four DNA nucleotides:
adenine and guanine are purines; cytosine and thymine are pyrimidines.

\section*{Acknowledgments}
One of the authors (H.Y.) would like to thank Drs. E. Starikov, D. Hennig and J. F. R. Archilla
for collaboration about the polaron models.
One of the authors (K.I.) would like to thank 
Kazuko Iguchi for her continuous
financial support and encouragement. 
We would like to thank Dr. Shu-ichi Kinosita for 
 sending us many relevant papers. 
 We would like to thank Dr. Eugene Starikov for very useful communications.



\begin{thebibliography}{00}
\bibitem{porath00} D. Porath, A.Bezryadin, S. de Varies and C. Dekker,
 Nature (London)  {\bf 403}, 635 (2000).

\bibitem{yoo01}
K. -H. Yoo {\it et al.}, Phys. Rev. Lett. {\bf 87}, 198102(2001).

\bibitem{porath04} For recent reviews, see example, 
 D. Porath, G. Curiberti and R. Difelice,
Topics in Current Chemistry, 1 (2004).

\bibitem{chakraborty07}
{\it Charge Migration in DNA: Perspectives from Physics, Chemistry, and Biology}, 
Ed. by Tapash Chakraborty, (Springer 2007). 

\bibitem{iguchi04} K. Iguchi, 
Int. J. Mod. Phys. B{\bf 18}, 1845(2004).

\bibitem{tran00} P. Tran, B. Alavi, and G. Gruner,
Phys. Rev. Lett. {\bf 85}, 1564(2000).

 \bibitem{yu01}
Z. G. Yu and Xueyu Song, 
Phys. Rev. Lett.  86, 6018(2001).

\bibitem{ladik99}
J. J. Ladik, {\it Quantum Theory of Polymers as Solids}, 
(Plenum Press, NY,1988); Phys. Rep.
313, 171 (1999).


\bibitem{berlin02}
Y.A.Berlin, A.L.Burin, M.A.Ratner,
Chem. Phys. {\bf 275}, 61-74(2002).

\bibitem{lin06}
Lin Xu {\it et al.}, 
Mol. Biol. Evol. {\bf 23}, 1107-1108(2006).


\bibitem{iguchi01} K. Iguchi,
 Int.J. Mod. Phys. B{\bf 11}, 2405(1997);
J. Phys. Soc. Jpn. {\bf 70}, 593(2001).

\bibitem{iguchi03} K. Iguchi,
 Int. J. Mod. Phys. B{\bf 17}, 2565(2003).

\bibitem{yamada04}
H. Yamada, Int. J. Mod. Phys. B {\bf 18}, 1697 (2004); Phys.
Lett. A {\bf 332}, 65 (2004).

\bibitem{yamada05}
H. Yamada,  E. B. Starikov, D. Hennig and J.F.R. Archilla,
Eur. Phys. J. E {\bf 17}, 149 (2005).

\bibitem{yamada07}
H. Yamada, E.B. Starikov, and D. Hennig, 
Euro. Phys. J. B, 59, 185-192 (2007). 

\bibitem{holste03} D. Holste, I. Grosse, S. Beirer, P. Schieg and H. Herzel,
Phys. Rev. E{\bf 67}, 061913(2003). 

\bibitem{isohata03} Y. Isohata and M. Hayashi,
J. Phys. Soc. Jpn. {\bf 72}, 735(2003). 


\bibitem{carpena02} P. Carpena,
  P. B. Galvan, P.Ch. Ivanov and H.E. Stanley,
Nature {\bf 418}, 955(2002); {\it ibid}, {\bf 84}, 764(2003).


\bibitem{grosse02} I. Grosse,  P. Bernaola-Galvan,  P. Carpena, 
 R. RomJan-RoldJan,  J. Oliver,  H.E. Stanley,
 Phys. Rev. E{\bf  65}  041905(2002).

\bibitem{krokhin09}
A. A. Krokhin et al.,
Phys. Rev. B 80 (2009) 085420.

 \bibitem{yamada04c} H. Yamada, 
{\it Slow Dynamics 2003} (AIP Conference Proceedings, 2004), 
Ed. by M. Tokuyama and I. Oppenheim,  773-774.

\bibitem{burmel69}
M. E. Burmel, D. D. Elley and V. Subramanyan,  
Ann. NY. Acad. Sci, {\bf 158}, 191 (1969).

\bibitem{fukui76}
K. Fukui, 
{\it Chemical Reactions and Electronic Orbitals},
(Maruzen, Tokyo, 1976), in Japanese.

\bibitem{huckel32}
E. H\"{u}ckel, Z. Phys. {\bf 76}, 628 (1932).

\bibitem{hoffman52}
R. Hoffman, J. Chem. Phys. {bf 39}, 1397 (1952). 

\bibitem{nagata65}
C. Nagata,
"Electronic Structures of DNA",
in {\it Quantum Chemistry II},
Lecture Series in Biophysics,
(Yoshioka Shoten, Kyoto, 1965), in Japanese;
{\it Introduction to Quantum Biology},
(Gakkai Shuppan Center, Tokyo, 1985), in Japanese.

\bibitem{burdett84}
J. K. Burdett, 
"From Bonds to Bands and Molecules to Solids",
Prog. Solid St. Chem. {\bf 15}, 173 (1984). 

\bibitem{sandorfy49}
C. Sandorfy, Bull. Soc. Chim. (France) {\bf 1949}, 615 (1949).

\bibitem{mulliken49}
R. S. Mulliken, C. A. Rieke, D. Orloff and H. Orloff, J. Chem. Phys. {\bf 17}, 1248 (1949).
R. S. Mulliken, J. Chem. Phys. {\bf 23}, 1841 (1955).

\bibitem{streitwieser61}
A. Streitwieser, Jr., 
{\it Molecular Orbital Theory for Organic Chemists\/}, (John Wiley \& Sons, New York, 1961), Chap. 5.


\bibitem{bruinsma00}
R. Bruinsma, G. Gruner, M. R. D'Orsogna, and J. Rudnick, 
Phys. Rev. Lett. 85, 4393?4396(2000).

\bibitem{roche03} S. Roche,
Phys. Rev. Lett. {\bf 91}, 108101(2003); S. Roche, D. Bicout, E. Macia and  E. Kats,
{\it ibid}, {\bf 91}, 22810(2003).

\bibitem{starikov02} E.B. Starikov,
J. Photochem. Photobiol. C {\bf 3}, 147(2002).

\bibitem{hennig02} D. Hennig,
Euro. Phys. J. B {\bf 30}, 211(2002): D. Hennig, J.F.R. Archilla and J. Agarwal,
Physica D {\bf 180}, 256(2003).

\bibitem{palmero04}  
F. Palmero, J.F.R. Archilla, D. Hennig and F.R. Romero, 
New J. Phys. {\bf 6}, 13(2004). 

\bibitem{chang04} C. Chang, A.H.C. Neto and A.R. Bishop, Chem. Phys. {\bf 115}, 4169(2004).

\bibitem{xray84} 
W. Saenger, {\it Principles of Nucleic Acid Structure}, (Springer, New York, 1984).


\bibitem{yamada01}
H. Yamada and T. Okabe, Phys. Rev. E {\bf 63}, 26203(2001).


\bibitem{crisanti93}
A. Crisanti, G. Paladin, and A. Vulpiani, {\it Products of Random
Matrices in Statistical Physics} (Springer-Verlag, Berlin, 1993),
and references therein.

 \bibitem{lifshits88} See, for example, L.M. Lifshits, S.A. Gredeskul and 
L.A. Pastur, {\it Introduction to the theory of Disordered Systems},
(Wiley, New York,1988).

\bibitem{imry02} Y. Imry and J. Imry,
{\it Introduction to Mesoscopic Physics 2nd edition},
(Oxford University Press, Oxford, 2002).

\bibitem{kats02} E.I. Kats and V.V. Lebedev,
JETP Letters {\bf 75}, 37(2002).

\bibitem{benjamin09}
P. Benjamin Woiczikowski, Toma Kuba, Rafael Gutierrez, 
Rodrigo A. Caetano,Gianaurelio Cuniberti, and Marcus Elstner,
J. Chem. Phys. 130, 215104 (2009). 


\bibitem{zalinge06}
H. van Zalinge, D. J. Schiffrin, A. D. Bates,
E. B. Starikov, W. Wenzel and R. J. Nichols, 
Angew. Chem. Int. Ed. {\bf 45}, 5499 (2006).

\bibitem{mandal06}
S. K. Mandal, 
Appl. Phys. Lett. {\bf 89}, 193102 (2006)


\bibitem{yamada99}
H.Yamada and K.S. Ikeda, Phys. Rev. E 59, 5214-5230 (1999).

\bibitem{yamada04a}
H. Yamada and K.S. Ikeda, Phys. Lett. A 328, 170-176(2004).

\bibitem{heinrichs02} J. Heinrichs, 
Phys. Rev. B {\bf 66}, 155434 (2002).


\bibitem{roche04}
S. Roche and E. Macia, 
Modern Physics Letters B, {\bf 18}, 847 (2004). 



\bibitem{su79}
W. P. Su, J. R. Schrieffer, and A. J. Heeger,
Phys. Rev. Lett. {\bf 42}, 1698 (1979);
Phys. Rev. {\bf B22}, 2099 (1980).

\bibitem{heeger88}
A. J. Heeger et. al., Rev. Mod. Phys. {\bf 60}, 781 (1988).


\bibitem{datta95}
S. Datta, 
{\it Electronic transport in mesoscopic systems}
 (Cambridge University Press 1995).

\bibitem{cornean04} 
Horia D. Cornean, Arne Jensen, V. Moldoveanu,
"A rigorous proof for the Landauer-Buttiker formula", 
DMF-2004-07-007, AALBORG-R-2004-10. 

\bibitem{macia06a}
E. Macia and S. Roche, 
Nanotechnology {\bf 17}, 3002 (2006).



\bibitem{mcfadden01}
J. McFadden, 
{\it Quantum Evolution}, (W W Norton and Co Inc, 2001).

\bibitem{lowdin63}
P.-O. L\"{o}wdin,
Rev. Mod. Phys. {\bf 35}, 724 (1963).  



 \bibitem{shih08}
C.T. Shih, S. Roche, R. Roemer, 
Phys. Rev. Lett. 100, 018105 (2008).


\bibitem{igarashi08}
A. Igarashi and H. Yamada, Phys. Rev. E 78, 026213-1-21 (2008).


\bibitem{gutierrez05}
R. Gutierrez, S. Mandal, and G. Cuniberti,
Phys. Rev. B 71, 235116 (2005).

\bibitem{gutierrez06}
R. Gutierrez, S. Mohapatra, H. Cohen, D. Porath, G. Cuniberti,
Phys. Rev. B 74, 235105 (2006).

\bibitem{macia06}
E. Macia,
Phys. Rev. B 74, 245105 (2006).

\bibitem{macia07}
E. Macia, 
Phys. Rev. B 75, 035130 (2007).


\bibitem{guo07}
Ai-Min Guo,
Phys. Rev. E 75, 061915 (2007).


\bibitem{malyshev07}
A. V. Malyshev,
Phys.Rev.Lett.98, 096801(2007).


\bibitem{ketabi09}
S. A. Ketabi and A A Fouladi,
PRAMANA J. Phys., 72, 1023-1036(2009).

\bibitem{gutierrez09}
R. Gutierrez and G. Cuniberti, R. Caetano, and T. Kubar,
arXiv:0910.0348v1 [cond-mat.soft] (2009).

\bibitem{bagci07}
V.M.K. Bagci, A.A. Krokhin,
Chaos, Solitons and Fractals 34, 104?111(2007).

\bibitem{molinari08}
Luca Guido Molinari,
Linear Algebra and its Applications 429 2221-2226(2008).

\bibitem{hori68}
J. Hori,
{\it Spectral Properties of Disorderd Chains and Lattices\/},
(Pergamon Press, Oxford, 1968).

\bibitem{schenk06}
O. Schenk, M. Bollhofer, R.A. Roemer,  
SIAM J. SCI. COMPUT. 28, 963-983(2006).


\bibitem{holstein59}
T. D. Holstein, 
Annals Phys. {\bf 8}, 325, 343 (1959);
L. A. Turkevich and T. D. Holstein,
Phys.Rev. {\bf B14}, 7474 (1987).



\bibitem{flach09}
S. Flach, D.O. Krimer, and Ch. Skokos, 
Phys. Rev. Lett. 102, 024101 (2009).

\bibitem{kopidakis08}
G. Kopidakis, S. Komineas, S. Flach, and S. Aubry, Phys.
Rev. Lett. 100, 084103 (2008).

\bibitem{pikovsky08}
A.S. Pikovsky and D.L. Shepelyansky, 
Phys. Rev. Lett. 100, 094101 (2008).

\bibitem{lahini08}
Y. Lahini {\it et al.}, 
Phys. Rev. Lett. 100, 013906(2008).

 \bibitem{larcher09}
M. Larcher,1 F. Dalfovo,1 and M. Modugno, 
Effects of interaction on the diffusion of atomic matter waves
in one-dimensional quasi-periodic potentials, 
arXiv:0909.1714v2.

\bibitem{garcia09}
I. Garcia-Mata and D.L. Shepelyansky,  
Phys. Rev. E 79, 026205 (2009). 

\bibitem{molinari03} L. Molinari, 
J. Phys. A: Math. Gen. {\bf 36}, 4081(2003).



\bibitem{aizawa84} Y. Aizawa, C. Murakami and T. Kohyama, 
Prog. Theor. Phys. Suppl. {\bf 79}, 96(1984);
Y. Aizawa, Chaos, Solitons and Fractals, {\bf 11}, 263(2000).

\bibitem{yamada91} H. Yamada, M. Goda and Y. Aizawa, 
J. Phys.:Condens. Matter {\bf 3}, 10043(1991).
 





















 \end{thebibliography}
\end{document}